\documentclass[%
reprint,
 amsmath,
 amssymb,
 aps,
 prb,
 superscriptaddress,
]{revtex4-2}

\usepackage{graphicx}
\usepackage[caption=false]{subfig}
\usepackage{dcolumn}
\usepackage{bm}
\newcommand{\ket}[1]{|{#1}\rangle}

\newcommand{\braket}[1]{\langle{#1}\rangle}
\usepackage[caption=false]{subfig}
\usepackage{soul}

\usepackage{mathtools}
\usepackage[T1]{fontenc}
\usepackage[utf8]{inputenc}
\usepackage{textcomp}
\usepackage[english]{babel}
\usepackage{lmodern}
\usepackage[babel]{microtype}

\DeclareUnicodeCharacter{2013}{--} 

\usepackage{xcolor}
\usepackage[colorlinks,
			linkcolor=black,
			citecolor=black,
			filecolor=black,
			urlcolor=blue!50!black,
			pdfusetitle]{hyperref}
\hypersetup{
	pdfauthor={Axel Fünfhaus, Marius Möller, Thilo Kopp, Roser Valent\'i},
	pdfsubject={Solid-State Physics},
	pdfkeywords={topology, magnetic field, interacting phases, Hofstadter model}
 }
\newcommand*\diff{\mathop{}\!\mathrm{d}}

\let\vec\boldsymbol

\def\ci{\mathrm{i}}
\DeclareMathOperator{\e}{e}

\begin{document}


\title{Topological Phase Transitions of Interacting Fermions in the Presence of a Commensurate Magnetic Flux}

\author{Axel Fünfhaus}
\email{fuenfhaus@itp.uni-frankfurt.de}
\affiliation{
 Institute of Theoretical Physics, Goethe University Frankfurt, Max-von-Laue-Straße 1, 60438 Frankfurt am Main, Germany
}
\author{Marius Möller}
\affiliation{
 Institute of Theoretical Physics, Goethe University Frankfurt, Max-von-Laue-Straße 1, 60438 Frankfurt am Main, Germany
}
\author{Thilo Kopp}
\affiliation{
 Center for Electronic Correlations and Magnetism, Institute of Physics, University of Augsburg, 86135 Augsburg, Germany
}
\author{Roser Valent\'i}
\affiliation{
 Institute of Theoretical Physics, Goethe University Frankfurt, Max-von-Laue-Straße 1, 60438 Frankfurt am Main, Germany
}

\date{\today}

\begin{abstract}
 Motivated by recently reported magnetic-field induced topological phases in ultracold atoms and correlated Moir\'{e} materials, we investigate topological phase transitions in a minimal model consisting of 
 interacting spinless fermions described by the Hofstadter model with Coulomb interaction on a square lattice. For interacting lattice Hamiltonians in the presence of a commensurate magnetic flux it has been demonstrated that the quantized Hall conductivity is constrained by a Lieb-Schultz-Mattis (LSM)-type theorem due to magnetic translation symmetry. In this work, we revisit the validity of the theorem for such models and establish that a topological phase transition from a topological to a trivial insulating phase can be realized but must be accompanied by spontaneous magnetic translation symmetry breaking caused by charge ordering of the spinless fermions. To support our findings, the topological phase diagram for varying interaction strength is mapped out numerically with exact diagonalization for different flux quantum ratios and band fillings using symmetry indicators. We discuss our results in the context of the LSM-type theorem.

\end{abstract}

\maketitle

\section{Introduction}\label{sec:Introduction}

Chern and topological insulators in condensed matter physics~\cite{Hal2, Kan} are closely linked to the discovery of the integer quantum Hall effect (IQHE)~\cite{Kli, Tho}. The problem of free electrons in a constant external magnetic field can be solved analytically and it leads to an understanding of the quantization of the Hall conductivity in terms of single-particle Landau levels~\cite{Lau, Halp2}. The characteristic high degeneracy and Chern numbers of the Landau levels are a result of the magnetic translation symmetry of electrons subject to an external magnetic field~\cite{Dan, Bro, Zak}. Similar symmetry constraints arising from magnetic translation symmetry also extend to non-interacting lattice Hamiltonians in the presence of a commensurate magnetic flux, such as the Hofstadter model on a square lattice~\cite{Hof}. Actually, it can be shown that for a gapped Hall insulator the Chern number $C$ of the occupied bands is constrained by a Diophantine equation of the form~\cite{Dan, Wan}
\begin{equation}\label{Eq:universal_relation}
    \e^{2\pi \ci (\frac{p}{q} C - \rho)} = 1,
\end{equation}
where $p/q$ is the flux quantum ratio per unit cell and $\rho$ is the number of particles per unit cell. In particular, the Chern number cannot be equal to zero if $\rho$ is not an integer, excluding a trivial band insulating phase. 

Recent theoretical predictions~\cite{Her, Fan, Gua, Her2, Zha2, Zha3} of additional topological invariants protected by translation symmetry in an external magnetic field have ignited a renewed interest in the Hofstadter model. Furthermore, there has been important experimental progress in cold atom gases~\cite{Aid, Gol} and twisted bilayer materials~\cite{Pol, Das} where the range of the flux quantum ratio can now go up to one elementary flux quantum $\Phi_0$ per unit cell. In particular, a bosonic fractional quantum Hall phase has been recently detected in the former~\cite{Lon}, and the formation of odd-denominator fractional quantum Hall states in graphene at high magnetic fields have been optically detected using excited Rydberg excitons in an adjacent transition metal dichalcogenide monolayer~\cite{Pop}.

For interacting systems, the classification of topological phases is more challenging than in the non-interacting case since the ground state is no longer given by a product state of single-particle Bloch wave functions. However, even without resorting to Bloch states in the Brillouin zone, it is possible to define a many-body Chern number using twisted boundary conditions which corresponds to adiabatic flux insertion and is thus related to the Hall conductivity~\cite{Niu}. This is now a well-established technique in mapping out topological phase diagrams of interacting Chern insulators~\cite{Var, Sha} and especially relevant for fractional Chern insulators, which are beyond a conventional mean-field description~\cite{Haf, Ger} and hence require a genuine many-body treatment. Due to the similarity of the Brillouin zone and the set of possible twisted boundary conditions, both forming a torus, it is a natural question to ask whether magnetic translation operators impose the same constraint, Eq.~(\ref{Eq:universal_relation}), on interacting systems.

An earlier and more well-known example of a filling-enforced constraint on quantum many-body systems is the (generalized) Lieb-Schultz-Mattis (LSM) theorem. Originally formulated for a one dimensional half-integer spin chain~\cite{Lie}, it has been generalized to higher dimensions~\cite{Hast} and many-body systems with conserved particle number on a periodic lattice~\cite{Osh}. Resting on the commutation relation between the ``large gauge transformation'' and the translation operator~\cite{Osh, Osh3}, the theorem determines whether a translation invariant Hamiltonian at a certain filling is allowed to have a non-degenerate ground state or not. Specifically, a unique ground state, separated by a gap from excited states, is prohibited for non-integer fillings. 

In the presence of an external magnetic field the algebra of translation operators changes to the magnetic translation algebra and integer fillings now correspond to $q\rho \in \mathbb{Z}$ due to an enlarged magnetic unit cell~\cite{Lu}. Even more importantly is the possibility to generalize Eq.~(\ref{Eq:universal_relation}) to genuine many-body systems~\cite{Avr, Lu, Mat, Kol, Man} and to exclude trivial band insulators at certain fillings. The proof is based on using within Laughlin's gauge argument the commutation relation of the large gauge transformation with the product of the magnetic translation operator and a flux insertion operator~\cite{Lu} in the thermodynamic limit. It may therefore be viewed as a corollary or variation of the generalized LSM theorem. This generalization can also be proven exactly for all finite-size systems on which the generalized LSM theorem applies~\cite{Mat}. We will therefore call it the LSM-{\it{type theorem}} in the following.

Crucially, the LSM-type theorem for the IQHE relies on a non-degenerate ground state such as a non-interacting band insulating state. In the case of ground state degeneracy however, which commonly takes place at a phase transition due to spontaneous symmetry breaking (SSB), the theorem's implications have to be reevaluated. 

In this work we scrutinize the physical implications of many-body magnetic translation invariance on the Hall conductivity in the light of SSB. In particular, the role of the specific choice of a given cluster, on which the validity of the LSM-type theorem and the mechanism of SSB depend in a subtle way, will be examined. For this purpose, we perform exact diagonalization (ED) calculations of the interacting Hofstadter model on a square lattice for a wide range of different band fillings and flux quantum ratios. Such calculations are useful since they  capture qualitative differences between different clusters. In the context of Hofstadter physics, ED allows us to study the effect of the discrete lattice (high flux quantum ratio) but also the parameter regime, where the single particle physics is approximately described by Landau levels (low flux quantum ratio). As a particular example for SSB, we investigate the translation symmetry breaking  charge-density-wave (CDW) state in the (long-range) Coulomb interaction regime. For the identification of topological phase transitions we make use of symmetry indicators~\cite{Mat}.

This paper is organized as follows: In Sec.~\ref{sec:first_section} we review basic properties of lattice Hamiltonians in the presence of a commensurate flux (in particular the Hofstadter model), the magnetic translation algebra and the constraint imposed by the LSM-type theorem on the Hall conductivity. In Sec.~\ref{sec:second_section} we provide an intuitive explanation backed by numerical calculations, how this constraint is circumvented by SSB. We use ED to discuss  the possibility of topological phase transitions  from Chern insulator to a checkerboard-pattern CDW phase for a half-integer flux quantum ratio. We expand the ED study by investigating all numerically feasible flux quantum ratios and integer band fillings of the interacting Hofstadter model on a square lattice in Sec.~\ref{sec:third_section}. We close our analysis by giving a thorough and rigorous account of the role of SSB in Sec.~\ref{sec:fourth_section} that addresses the remaining formal and technical questions of the applicability of the LSM-type theorem in integer as well as fractional quantum Hall systems. Finally, in Sec.~\ref{sec:Conclusion} we conclude our findings and discuss future directions.

\section{Lattice Hamiltonians in Commensurate Magnetic Flux}\label{sec:first_section}

\subsection{Hofstadter model}

To understand how a tight-binding Hamiltonian is affected by an external magnetic field, one can consider a particle of charge $-e$ moving around a surface that is pierced by a flux $\Phi$. The wave function obtains an Aharonov-Bohm phase of $-2\pi \Phi / \Phi_0$, where $\Phi_0 = hc/e$ is the magnetic flux quantum, which can be seen by considering the change in the Lagrange function $L \to L - e \dot{\vec{r}}\vec{A}/c$ upon adding a flux $\Phi$ which leads to the phase
\begin{equation}\label{Peierls}
    \e^{-\ci \frac{e}{\hbar c} \int \diff \vec{r} \vec{A}(\vec{r})}
\end{equation}
due to the change in the action $S[\vec{r}] = \int_{0}^{t} \diff t' L[\dot{\vec{r}}, \vec{r}]$ in the path weight $\e^{\ci S[\vec{r}]/\hbar}$. On a single-orbital lattice model an external magnetic field can be implemented by including so-called Peierls phases to the hopping matrix elements for spinless fermions~\cite{Hof, Lut}. Such phases can be interpreted as Aharonov-Bohm phases originating from a magnetic flux piercing the different plaquettes of the given lattice:
\begin{equation}
    c_{\vec{R}'}^{\dagger} c_{\vec{R}}^{\phantom{\dagger}} \to c_{\vec{R}'}^{\dagger} c_{\vec{R}}^{\phantom{\dagger}} \e^{-\ci \frac{e}{\hbar c} \int_{\vec{R}}^{\vec{R}'} \vec{A}(\vec{r}') \diff \vec{r}'} =: c_{\vec{R}'}^{\dagger} c_{\vec{R}}^{\phantom{\dagger}} \e^{- \ci \phi_{\vec{R}', \vec{R}}}.
\end{equation}
The flux through a set of plaquettes enclosed by the path $\vec{R}_1, \vec{R}_2, \dots, \vec{R}_n, \vec{R}_1$ is given by the sum $\phi_{\vec{R}_2, \vec{R}_1} + \phi_{\vec{R}_3, \vec{R}_2} + \dots + \phi_{\vec{R}_1, \vec{R}_n}$. A spin-polarized (or equivalently spinless) single-orbital nearest-neighbor (NN) tight-binding Hamiltonian on a square lattice with a flux quantum ratio $\Phi/\Phi_0 = \varphi$ per plaquette becomes
\begin{equation}\label{Eq:Hofstadter_infinite_plane}
    \begin{aligned}
    \hat{H} = & -t \displaystyle\sum_{m,n} c_{m+1, n}^{\dagger} c_{m,n}^{\phantom{\dagger}} \\
    & - t \displaystyle\sum_{m,n} \e^{-2\pi \ci \varphi m} c_{m,n+1}^{\dagger} c_{m,n}^{\phantom{\dagger}} + \text{h.c.},
    \end{aligned}
\end{equation}
which is the Hofstadter model in the Landau gauge~\cite{Hof}. Different choices for the phases $\phi_{\vec{R}', \vec{R}}$ are equivalent up to a unitary gauge transformation $c_{m,n}^{\dagger} \to e^{i\lambda_{m,n}} c_{m,n}^{\dagger}$~\cite{Kob} as long as the flux piercing each of the plaquettes is the same~\cite{Ste}. The Chern numbers of the band spectrum of Eq.~(\ref{Eq:Hofstadter_infinite_plane}) are uniquely determined by a specific Diophantine equation for all possible $\varphi$~\cite{Fue, Tho}.

\subsection{Magnetic translation algebra}

Since hopping amplitudes and fluxes of the Hamiltonian in Eq.~(\ref{Eq:Hofstadter_infinite_plane}) are translation invariant, one can identify two unitary translation operators in $\vec{x}$ and $\vec{y}$ direction defined by
\begin{equation}\label{Eq:magn_trans_second_quant}
\begin{aligned}[b]
    \hat{T}_{\vec{x}}^{M} c_{m,n}^{\dagger} \left(  \hat{T}_{\vec{x}}^{M}\right)^{\dagger} = & \e^{-2\pi \ci \varphi n} c_{m+1,n}^{\dagger}\\
    \hat{T}_{\vec{y}}^{M} c_{m,n}^{\dagger} \left(  \hat{T}_{\vec{y}}^{M}\right)^{\dagger} = & c_{m,n+1}^{\dagger}.
\end{aligned}
\end{equation}
These operators, consisting of the action of a symmetry operation in real space times a unitary and space dependent gauge transformation are commonly denoted magnetic translation operators~\cite{Bro, Zak}. They generally do not commute with each other. Acting on a Hilbert space of $N_e$ particles they satisfy~\footnote{Eq.~(\ref{Eq:magn_trans_b}) indicates that magnetic translation operators transform as a projective representation. It is deduced in the continuum from the Baker-Campbell-Hausdorff formula, whereas in a discrete Hilbert space it simply serves as the definition of $\hat{T}_{\vec{a}}^{M}$ for arbitrary Bravais lattice vectors $\vec{a}$. In particular it should be noted that for $\varphi = 2\pi$, Eq.~(\ref{Eq:magn_trans_b}) does not realize a non-trivial projective rep, as the magnetic translation operators are abelian according to Eq.~(\ref{Eq:magn_trans_a}).}
\begin{subequations}\label{Eq:magn_trans}
    \begin{alignat}{2}
        \hat{T}_{\vec{x}}^{M} \hat{T}_{\vec{y}}^{M} = & \e^{-2 \pi \ci \varphi N_e} \hat{T}_{\vec{y}}^{M} \hat{T}_{\vec{x}}^{M} \label{Eq:magn_trans_a} \\
        \hat{T}_{\vec{x}}^{M} \hat{T}_{\vec{y}}^{M} = & \e^{-\pi \ci \varphi N_e} \hat{T}_{\vec{x} + \vec{y}}^{M}, \label{Eq:magn_trans_b}
    \end{alignat}
\end{subequations}
which is referred to as the magnetic translation algebra or Girvin-MacDonald-Platzman algebra~\cite{Ber}. According to Eq.~(\ref{Eq:magn_trans}) these operators form a higher-dimensional projective representation~\cite{Bac, Bac2, Yan} for $\varphi N_e \notin \mathbb{Z}$, in contrast to the usual one-dimensional irreducible representations for the abelian translation group. A consequence is a symmetry-protected degeneracy of the energy spectrum: for example, the macroscopic degeneracy of single-particle Landau levels stems from the magnetic translation symmetry of free particles~\cite{Bro}.

For a non-interacting lattice Hamiltonian such as the Hofstadter model and for a flux quantum ratio 
\begin{equation}
    \varphi = \frac{p}{q},
\end{equation}
where in the following $p$ and $q$ will always be chosen to be coprime, the single-particle band structure obtains a symmetry protected $q$-fold degeneracy. This can be seen as follows: Since $(\hat{T}_{\vec{x}}^{M})^{q}, \hat{T}_{\vec{y}}^{M}$ are commuting translation operators according to Eq.~(\ref{Eq:magn_trans_a}) for $N_e = 1$, they can be used to define Bloch states $\ket{k_x, k_y}$ with eigenvalues $\e^{-\ci q k_x}, \e^{-\ci k_y}$ (where we have set the lattice constant to 1). Then, since
\begin{equation}\label{Eq:magn_transl_degeneracy}
    \hat{T}_{\vec{y}}^{M} \left( \hat{T}_{\vec{x}}^{M} \ket{k_x, k_y} \right) = \e^{-\ci (k_y - 2\pi \varphi)} \left( \hat{T}_{\vec{x}}^{M} \ket{k_x, k_y} \right)
\end{equation}
we obtain that $\hat{T}_{\vec{x}}^{M} \ket{k_x, k_y}$ is a new eigenstate of $\hat{T}_{\vec{y}}^{M}$, hence 
\begin{equation}
    \hat{T}_{\vec{x}}^{M} \ket{k_x, k_y} \propto \ket{k_x, k_y - 2\pi \varphi}.
\end{equation}
As a consequence $\ket{k_x, k_y}, \dots, \ket{k_x, k_y - 2\pi \varphi (q-1)}$ span a $q$-dimensional irreducible projective representation space with degenerate energy eigenvalues. This also explains why according to Eq.~(\ref{Eq:universal_relation}), the Chern number can only change modulo $q$ because the topological phase transition of a band is accompanied with the simultaneous closure of $q$ band gaps. A complete derivation of Eq.~(\ref{Eq:universal_relation}) is given in Appendix~\ref{App:zero_appendix}.

\begin{figure}[ht]
 	\centering
 	\includegraphics[width=1.0\linewidth]{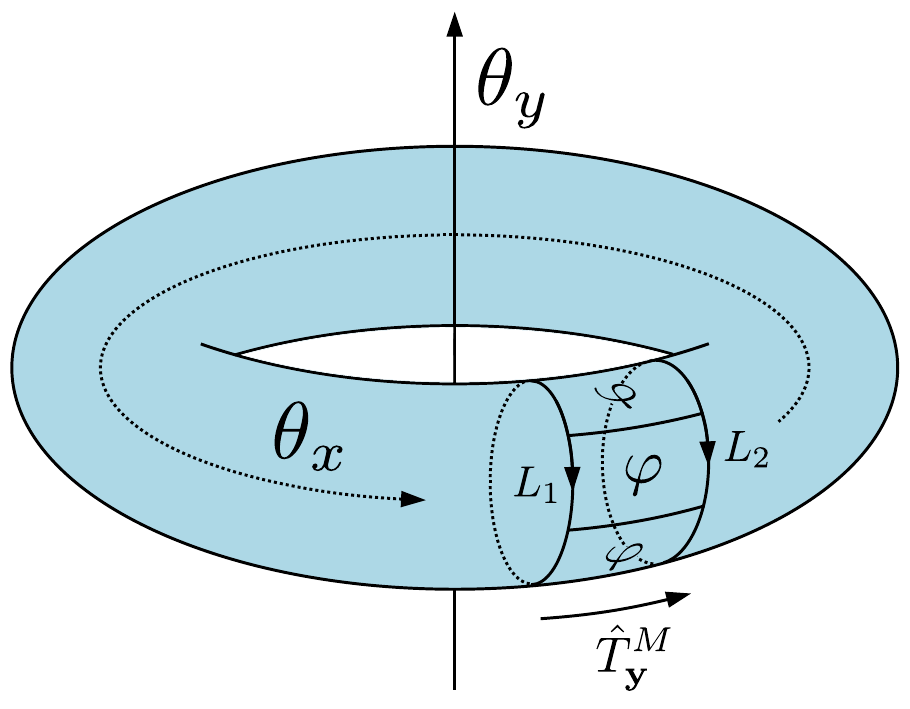}
 	\caption{Toroidal lattice geometry of a finite cluster in real space with two fluxes corresponding to twisted boundary conditions $\theta_x$ and $\theta_y$ through the non-contractable loops of the torus. A translation in $\vec{y}$-direction relates the loops $L_1$ and $L_2$ to each other. Their Aharonov-Bohm phases differ by the flux $2\pi \varphi N_x$ through the surface between the two loops.}
 	\label{fig:torus_fluxes}
 \end{figure}

\begin{figure}[ht]
 	\centering
 	\includegraphics[width=1.0\linewidth]{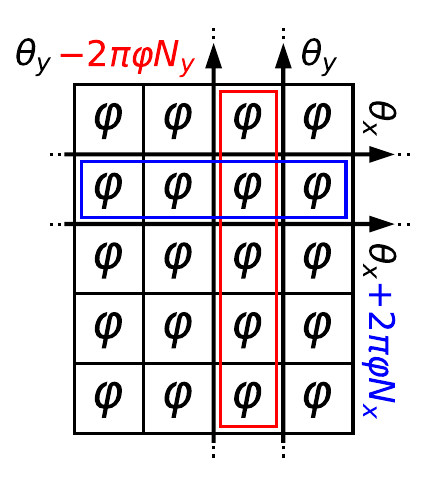}
 	\caption{Finite size $N_x \times N_y$ cluster. The arrows represent different non-contractible loops of the torus.}
 	\label{fig:lattice_twisted_boundaries}
 \end{figure}

For interacting systems ED calculations have to be restricted to finite systems. The infinite plane becomes a cluster with its boundaries being glued together forming a torus. Besides piercing each plaquette with a flux, it is now possible to insert fluxes through the non-contractible loops of the torus~\cite{Niu, Ste}, see Fig.~\ref{fig:torus_fluxes}. They can be varied by imposing twisted boundaries onto the system, which keep the fluxes through the individual plaquettes invariant. The Hamiltonian in Eq.~(\ref{Eq:Hofstadter_infinite_plane}) becomes
\begin{equation}\label{Eq:Hofstadter_torus}
    \begin{aligned}[b]
    \hat{H}(\theta_x, \theta_y) = & -t \displaystyle\sum_{m = 1}^{N_x} \displaystyle\sum_{n = 1}^{N_y} \e^{- \ci \theta_x \delta_{m, N_x}} c_{m+1, n}^{\dagger} c_{m,n}^{\phantom{\dagger}} \\
    & - t \displaystyle\sum_{m = 1}^{N_x} \displaystyle\sum_{n = 1}^{N_y} \e^{-2\pi \ci \varphi m} \e^{- \ci \theta_y \delta_{n, N_y}}c_{m,n+1}^{\dagger} c_{m,n}^{\phantom{\dagger}} \\
    & + \text{h.c.},
    \end{aligned}
\end{equation}
with twisted boundary $\theta_x$ in $\vec{x}$-direction and $\theta_y$ in $\vec{y}$-direction. $N_{x/y}$ is the number of sites in $\vec{x}/\vec{y}$-direction. We will always consider $N_x / q \in \mathbb{N}$, unless stated otherwise. This choice automatically satisfies the quantization condition that the total flux through all of the plaquettes can only be zero modulo $2\pi$, as sources of magnetic flux may only emit integer multiples of $\Phi_0$~\cite{Bro, God}. \vspace{-0.2cm}
   \vspace{-0.1cm}
Importantly, a tight-binding Hamiltonian of the form of Eq.~(\ref{Eq:Hofstadter_torus}) is not translation invariant for arbitrary $N_x, N_y$. Let the Aharonov-Bohm phase around a non-contractable loop $L_2$ in $\vec{x}$-direction for fixed $n$ be
\begin{equation}
    \displaystyle\sum_{m =1}^{N_x} \phi_{(m+1, n), (m,n)} = \theta_x.
\end{equation}
Around a similar loop $L_1$ for fixed $n-1$ we identify
\begin{equation}
    \begin{aligned}[b]\label{Eq:different_loops}
       & \displaystyle\sum_{m = 1}^{N_x} \phi_{(m+1, n-1),(m, n-1)} = \displaystyle\sum_{m = 1}^{N_x} \left( \phi_{(m+1, n-1),(m, n-1)} \right. \\
        + & \left. \phi_{(m+1, n),(m+1, n-1)} + \phi_{(m, n),(m+1, n)} + \phi_{(m, n-1),(m, n)} \right) \\
       + & \displaystyle\sum_{m=1}^{N_x} \phi_{(m+1,n), (m,n)} = 2\pi \varphi N_x + \theta_x,
    \end{aligned}
\end{equation}
where we have used $\phi_{\vec{R}', \vec{R}} = - \phi_{\vec{R}, \vec{R}'}$. Note that the first four terms in between the equal signs generate a loop around a plaquette with flux $\varphi$. We see that generically the magnetic field piercing the torus also contributes to the phase, see Fig.~\ref{fig:torus_fluxes}. The loop defined in Eq.~(\ref{Eq:different_loops}) is only equal to $\theta_x$ modulo $2\pi$, if $N_x$ is an integer multiple of $q$. For $N_y$ an analogous condition for closed loops with fixed $m$ holds. Consequently, under translation the Hamiltonian transforms as~\footnote{Here we have omitted the dependency of the translation operators on the twisted boundary condition for simplicity, which does not influence the validity of Eq.~(\ref{Eq:universal_relations_many_body})~\cite{Mat}. In~\cite{Lu} it is pointed out that the definition of the magnetic translation operator as it appears in Eq.~(\ref{Eq:magn_trans_second_quant}) cannot be used anymore if the Hamiltonian does not commute with the translation operator. Instead, a gauge of $\hat{H}(\vec{\theta})$ for all $\vec{\theta}$ needs to be defined so that the magnetic translation operator maps the Hamiltonians of this gauge to each other.}
\begin{equation}\label{Eq.many-body_magn_trans_algebra}
    \begin{aligned}[b]
        \hat{T}_{\vec{x}}^{M} \hat{H}(\theta_x, \theta_y) \left( \hat{T}_{\vec{x}}^{M} \right)^{\dagger} = & \hat{H}(\theta_x, \theta_y - 2\pi \varphi N_y) \\
        \hat{T}_{\vec{y}}^{M} \hat{H}(\theta_x, \theta_y) \left( \hat{T}_{\vec{y}}^{M} \right)^{\dagger} = & \hat{H}(\theta_x + 2\pi \varphi N_x, \theta_y),
    \end{aligned}
\end{equation}
see Fig.~\ref{fig:lattice_twisted_boundaries}, so the system is only translation invariant if $N_x$ and $N_y$ are integer multiples of $q$~\cite{Bro, Zak}.

\subsection{Constraints imposed by the LSM-type theorem on the Hall conductivity}

 It is possible to define a many-body Chern number of the ground state $\ket{\Psi(\theta_x, \theta_y)}$ with the Berry connection $\vec A = \ci \braket{\Psi(\theta_x, \theta_y)|\vec{\nabla}_{\vec{\theta}}|\Psi(\theta_x, \theta_y)}$), which is related to the Hall-conductivity by Laughlin's gauge invariance argument~\cite{Niu}. This Chern number is also well defined upon adding arbitrary interactions to the Hamiltonian as long as an energy gap separates the ground state manifold from excitations for all twist angles. A natural question that arises is whether it is possible to generalize Eq.~(\ref{Eq:universal_relation}) to many-body systems. In the context of the generalized LSM theorem it can be shown that~\cite{Lu, Mat, Avr}
\begin{equation}\label{Eq:universal_relations_many_body}
     \e^{2\pi \ci (\frac{p}{q} C_{\text{many-body}} - \rho)} = 1
 \end{equation}
 with the many-body Chern number $C_{\text{many-body}}$. This holds provided that a given twist angle combination $(\theta_x, \theta_y)$ is mapped to $q$ different twist angles under translation and the system has a unique ground state for each $\vec \theta$ (see Appendix~\ref{App:zero_appendix}). In the following we will refer to Eq.~(\ref{Eq:universal_relations_many_body}) as the LSM-type theorem. Clusters, where the LSM-type theorem strictly holds are those where the Hamiltonian $\hat{H}(\theta_x, \theta_y)$ is transformed into $q$ different $\hat{H}(\theta_x', \theta_y')$ under translation. We will call these clusters LSM clusters. On the other hand, if translation maps $\vec{\theta}$ onto itself so that the Hamiltonian commutes with the translation operator, we will speak of a non-LSM cluster. Naturally, every cluster containing $c_x q \times c_y q$ lattice sites ($c_x, c_y \in \mathbb{N}$) is a non-LSM cluster as a consequence of Eq.~(\ref{Eq.many-body_magn_trans_algebra}), but can be made into an LSM cluster by adding one row either in $\vec{x}$- or $\vec{y}$-direction. For example, for $\varphi = 1/2$ the cluster displayed in Fig.~\ref{fig:lattice_twisted_boundaries} with $N_x = 4$ and $N_y = 5$ would be an LSM cluster, but it could be changed into a non-LSM cluster by either adding or removing one row in $\vec{y}$-direction. It is important to realize that Eq.~(\ref{Eq:universal_relations_many_body}) cannot be proven for non-LSM clusters, even though it is to be expected that observables such as the Hall conductivity should not change in a significant way by adding a single row in $\vec{x}$- or $\vec{y}$-direction when approaching the thermodynamic limit.

Interaction terms, such as nearest-neighbor or long-range Coulomb interaction, do not affect the translation invariance of the Hamiltonian. Therefore, one may not anticipate interaction-driven topological phase transitions from a non-trivial Chern insulating to a trivial insulating phase because of the LSM-type theorem Eq.~(\ref{Eq:universal_relations_many_body}). On the other hand, sufficiently large Coulomb interaction is expected to induce localized charge ordered phases such as Wigner-crystal or CDW states~\cite{Kup, Pan, Var} that have zero Hall conductivity. In Sec.~\ref{sec:second_section} and~\ref{sec:third_section} we will provide numerical evidence for the intuitive claim that long-range interaction drives the system into a topologically trivial CDW state, before rigorously discussing in Sec.~\ref{sec:fourth_section} how SSB of translation invariance leads to a breakdown of the constraint on the Hall conductivity given by the LSM-type theorem.

\section{Topological Phase Diagram of the interacting Hofstadter Model for Half-Integer Flux Quanta}\label{sec:second_section}

  \begin{figure}[ht]
 	\centering
 	\includegraphics[width=1.0\linewidth]{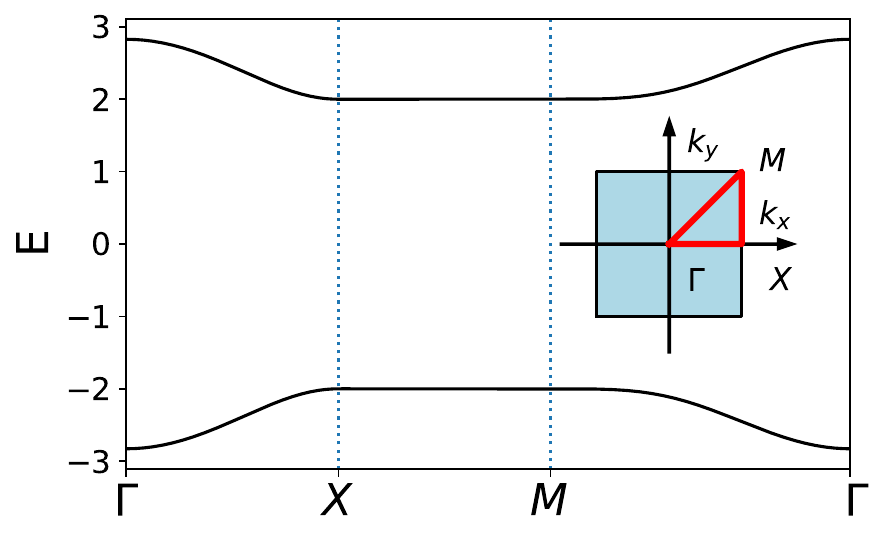}
 	\caption{Band spectrum of the $t$-$t'$--Hofstadter model for $\varphi = 1/2$ along the high symmetry path $\Gamma (0,0) \to X(\pi/q, 0) \to M(\pi/q, \pi/q) \to \Gamma(0,0)$ of the magnetic wallpaper group $p4m'm'$. The choice of hopping parameters of $t = 1 = 2t'$ results in a relatively flat band structure, mimicking the physics of Landau levels. The single-particle energy gap is exactly equal to $4t$.}
 	\label{fig:hofstadter_spectrum_1_2}
 \end{figure}

The non-interacting nearest-neighbor (NN) Hofstadter model with $\varphi = 1/2$ has symmetry-protected Dirac cones due to time reversal invariance (TRI)~\cite{Hat, Wen} because every closed loop contains an integer multiple of $\pi$ fluxes. 

We include next-nearest-neighbor (NNN) hoppings to break TRI,
\begin{equation}\label{Eq:nnn_hofstadter}
\begin{aligned}[b]
    \hat{H} = & \displaystyle\sum_{m = 1}^{N_x} \displaystyle\sum_{n = 1}^{N_y} \left\lbrace -t \e^{-\ci \theta_{x} \delta_{m, N_x}} c_{m+1, n}^{\dagger} c_{m,n}^{\phantom{\dagger}} \right.\\
    & - t \e^{-\ci \theta_{y} \delta_{n, N_y}} \e^{- \pi m \ci} c_{m,n+1}^{\dagger} c_{m,n}^{\phantom{\dagger}} \\
    & - t' \e^{-\ci \theta_{x} \delta_{m, N_x}} \e^{-\ci \theta_{y} \delta_{n, N_y}} \e^{-\pi (m + 1/2) \ci} c_{m+1, n+1}^{\dagger} c_{m,n}^{\phantom{\dagger}} \\
    & \left. - t' \e^{-\ci \theta_{x} \delta_{m, N_x}} \e^{\ci \theta_{y} \delta_{1, N_x}} \e^{\pi(m + 1/2)\ci} c_{m+1, n-1}^{\dagger} c_{m,n}^{\phantom{\dagger}} \right\rbrace\\
    & + \text{h.c.}
\end{aligned}
\end{equation}
where the Peierls phases are again determined with Eq.~(\ref{Peierls}) in Landau gauge. We set $t = 1 = 2t'$. The band structure and Brillouin zone are shown in Fig.~\ref{fig:hofstadter_spectrum_1_2}. The energetically lower band of the single-particle band structure has a Chern number of $C = 1$, in agreement with Eq.~(\ref{Eq:universal_relation}). We include nearest-neighbor Coulomb interactions: 
\begin{equation}\label{Eq:short_range_coulomb}
    \hat{H}_{\text{int}}(V) = V \displaystyle\sum_{m = 1}^{N_x} \displaystyle\sum_{n = 1}^{N_y} (\hat{n}_{m,n} \hat{n}_{m+1,n} + \hat{n}_{m,n} \hat{n}_{m,n+1}).
\end{equation}
that drive the Chern insulating phase into a trivial insulating phase.  In the limit $V \gg t$ we expect at half-filling a CDW-phase. If we imagine the cluster to be in a checkerboard structure, then the two ground state wave functions will be a product state of all particles being localized at ``black'' or ``white'' squares, instead of the delocalized Wannier functions of a Chern insulator. The charge order is indicated by the static structure factor
\begin{equation}\label{Eq:cdw_structure_factor}
    S(\vec{K}) = \frac{1}{(N_x N_y)^2} \displaystyle\sum_{\vec{r}, \vec{r}'} \e^{-\ci \vec{K}(\vec{r} - \vec{r}')} \braket{\hat{n}_{\vec{r}} \hat{n}_{\vec{r}'}}
\end{equation}
which for a checkerboard pattern will have a peak at $\vec{K} = (\pi, \pi)$, whereas $S(\vec{K})$ approaches zero at other $\vec{K}$. In addition, quantum phase transitions of interacting systems are indicated by the fidelity metric~\cite{Alb, Che, Cam}
\begin{equation}\label{Eq:fidelity_metric_overlap}
    g(V, \delta V) = \frac{2}{N_x N_y} \frac{1 - |\braket{\Psi_{0}(V)|\Psi_{0}(V + \delta V)}|}{(\delta V)^2},
\end{equation}
where in our case $\ket{\Psi_0(V)}$ is the ground state of $\hat{H} + \hat{H}_{\text{int}}(V)$. In the vicinity of a critical point one expects a peak of the fidelity metric, irrespective of whether a possible order parameter is commensurable with the chosen cluster. Instead of the fidelity metric given in Eq.~(\ref{Eq:fidelity_metric_overlap}) we evaluate the negative second derivative of the ground state energy with respect to $V$, which is closely related to the fidelity metric~\cite{Alb, Che} and provides qualitatively the same information, but is numerically less costly to evaluate:
\begin{equation}
   \chi_{E}(V) = - \frac{d^2 E_{0}(V)}{d V^2}.
\end{equation}
In particular, the peak of $\chi_{E}(V)$ also indicates a phase transition.

The topology of the ground state(s) can also change concomitantly with a CDW phase transition~\cite{Var}. Whereas the phase transition to the CDW-phase can occur due to an avoided crossing of ground states with a different order parameter, a topological phase transition must have a level crossing at some twist angle $\theta$. The Chern number is expected to change from 1 in the Chern insulating phase to 0 in the CDW-phase at high $V$. It is then possible to focus on the rotation eigenvalues at the three high symmetry points $(\theta_x, \theta_y) = (0,0), (\pi, \pi)$ and $(0, \pi)$ (which by $C_4$ symmetry is related to the point at $(\pi, 0)$), where the level-crossing must occur~\cite{Var}. This is the case because the Chern number is given modulo 4 by the eigenvalues of the rotation operators~\cite{Mat}:
\begin{equation}\label{Eq:symmetry_formula_c_4}
    \e^{- \frac{\pi \ci C}{2}} = W_{C_4^{+}}^{(0,0)} W_{C_4^{+}}^{(\pi,\pi)} W_{C^{\phantom{+}}_2}^{(0,\pi)}
\end{equation}
with
\begin{equation}
    W_{C_n}^{(\theta_x, \theta_y)} = \braket{\Psi(\theta_x, \theta_y)|C_n|\Psi(\theta_x, \theta_y)}
\end{equation}
and the rotation operators (for arbitrary flux $\varphi$) defined by
\begin{equation}
    \begin{aligned}
        C_4^{+} c_{m,n}^{\dagger} \left( C_4^{+} \right)^{\dagger} = \e^{\ci \left( 2\pi \varphi m n - 2\pi \varphi (N_x + 1)m + \frac{\pi \varphi}{2} (N_x + 1)^2 \right)} \\ 
      c_{-n + N_x + 1, m}^{\dagger}
    \end{aligned}
\end{equation}
(with $N_x = N_y$) and
\begin{equation}
    \begin{aligned}
       C_2 c_{m,n}^{\dagger} C_2^{\dagger} & = \e^{\ci \left( 2\pi \varphi n (N_x + 1) - \pi \varphi (N_x + 1)(N_y + 1) \right)} \\  
    &  \qquad\qquad\qquad\qquad c_{-m + N_x + 1, -n + N_y + 1}^{\dagger}.
    \end{aligned}
\end{equation}
 Note that in this form $(C_4^{+})^4 = \openone = C_2^2$. A gauge transformation has to be added to the pure rotation because we do not use a gauge with a rotation-symmetric vector potential for the Hamiltonian: it cannot be defined for certain cluster geometries such as $q \times q$ clusters. For the following calculations for non-LSM clusters the Hamiltonian $\hat{H}(\vec{\theta})$, Eq.~(\ref{Eq:nnn_hofstadter}), will be considered, however exclusively in the regime of twist angles $\vec{\theta}$ where the change of the symmetry eigenvalue and hence the level crossing is identified.
 
 \begin{figure}[ht]
 	\centering
 	\includegraphics[width=1.0\linewidth]{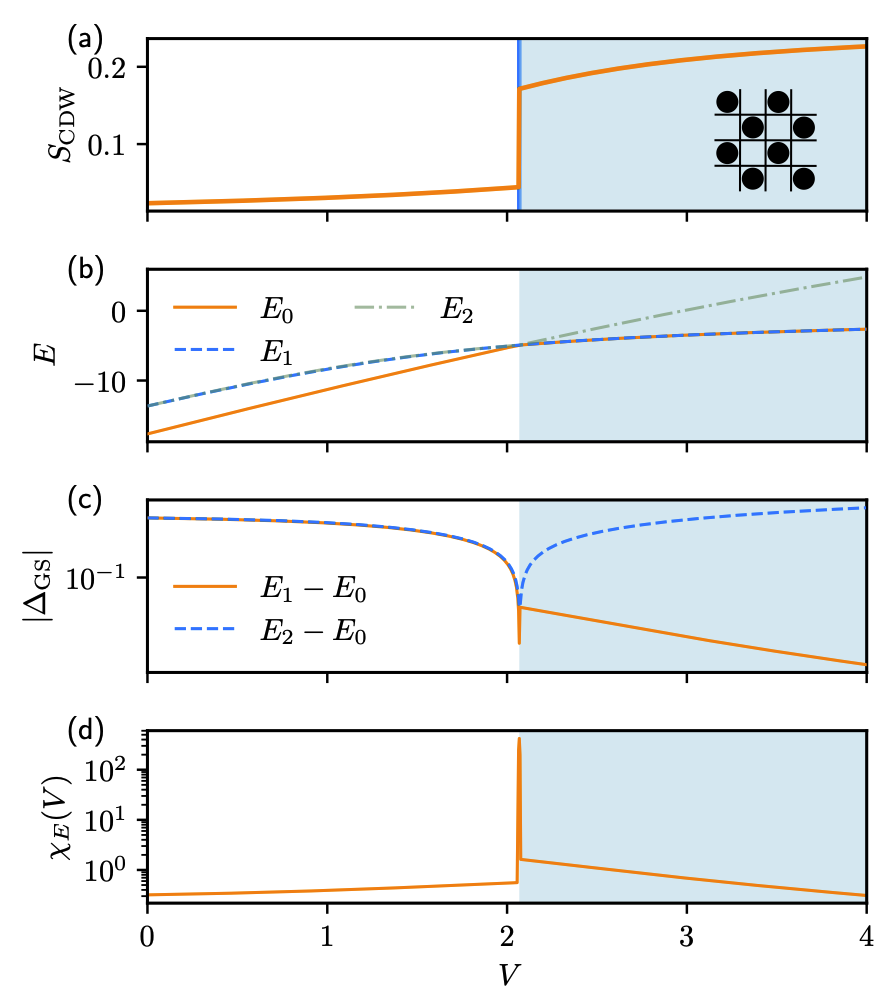}
 	\caption{ED of a non-LSM $4 \times 4$ cluster. (a) Static structure factor $S_{\text{CDW}}$ at $\vec{K} = (\pi, \pi)$ corresponding to a checkerboard pattern. (b) Energy of the ground state $E_0$ and first two excited states ($E_1, E_2$). (c) Energy gap between $E_0$ and ($E_1, E_2$). (d) Second derivative of $E_0$ as a function of the interaction strength $V/t$. The interaction strength is varied in steps $\delta V = 4/1000$. The IQHE and CDW phases are displayed in white and blue, respectively.}
 	\label{fig:CDW_Hofstadter_4_4_1_2}
 \end{figure}

We first display calculations for $N_x \times N_y$ clusters of the Hamiltonian in Eq.~(\ref{Eq:nnn_hofstadter}), where both $N_x$ and $N_y$ are multiples of $q = 2$. These clusters are non-LSM clusters, so they are technically not constrained by Eq.~(\ref{Eq:universal_relations_many_body}) and the $\vec{\theta}$-dependence of an eigenstate can yield a trivial Chern number. Note that these clusters are also compatible with the checkerboard pattern CDW. In Fig.~\ref{fig:CDW_Hofstadter_4_4_1_2} we display calculations of a $C_4$-symmetric cluster with $4 \times 4$ sites at $\vec{\theta} = (0,0)$ using an implicit restarted Arnoldi method implemented in Arpackpp. The phase transition from the Hall insulating to the CDW-state is clearly indicated by a jump in the CDW structure factor and a peak in $\chi_{E}(V)$ at $V = 2.068 \pm 0.04$. In the CDW-phase, the system has a quasi-degenerate (= degenerate up to some finite size splitting) twofold ground state, corresponding to the two possible charge-density-wave patterns with a small finite size gap. The symmetry eigenvalue $W_{C_4}^{(0,0)}$ of the lower energy ground state changes from $-\ci$ to $1$, which indicates a Chern-trivial phase of the ground state according to Eq.~(\ref{Eq:symmetry_formula_c_4}). We verified the absence of accidental crossings for different twist angles in the CDW phase, which ensures that the many-body Chern number of the lower energy ground state is well defined.


To take into account finite-size effects, calculations on a $4 \times 6$ cluster have also been performed. Note that this cluster breaks $C_4$ rotation symmetry, so instead of Eq.~(\ref{Eq:symmetry_formula_c_4}) the many-body Chern number is determined with
\begin{equation}\label{Eq:symmetry_formula_c_2}
    \e^{\pi \ci C} = W_{C_2}^{(0,0)} W_{C_2}^{(0,\pi)} W_{C_2}^{(\pi,0)} W_{C_2}^{(\pi,\pi)}.
\end{equation}
There is no qualitative difference in the results of these calculations compared to those shown in Fig.~\ref{fig:CDW_Hofstadter_4_4_1_2}. The phase transition occurs at $V = 2.078 \pm 0.003$ and is again accompanied by a change of the Chern number of the ground state, given by the symmetry indicator in Eq.~(\ref{Eq:symmetry_formula_c_2}).

\begin{figure}[htp]
  \includegraphics[clip,width=\columnwidth]{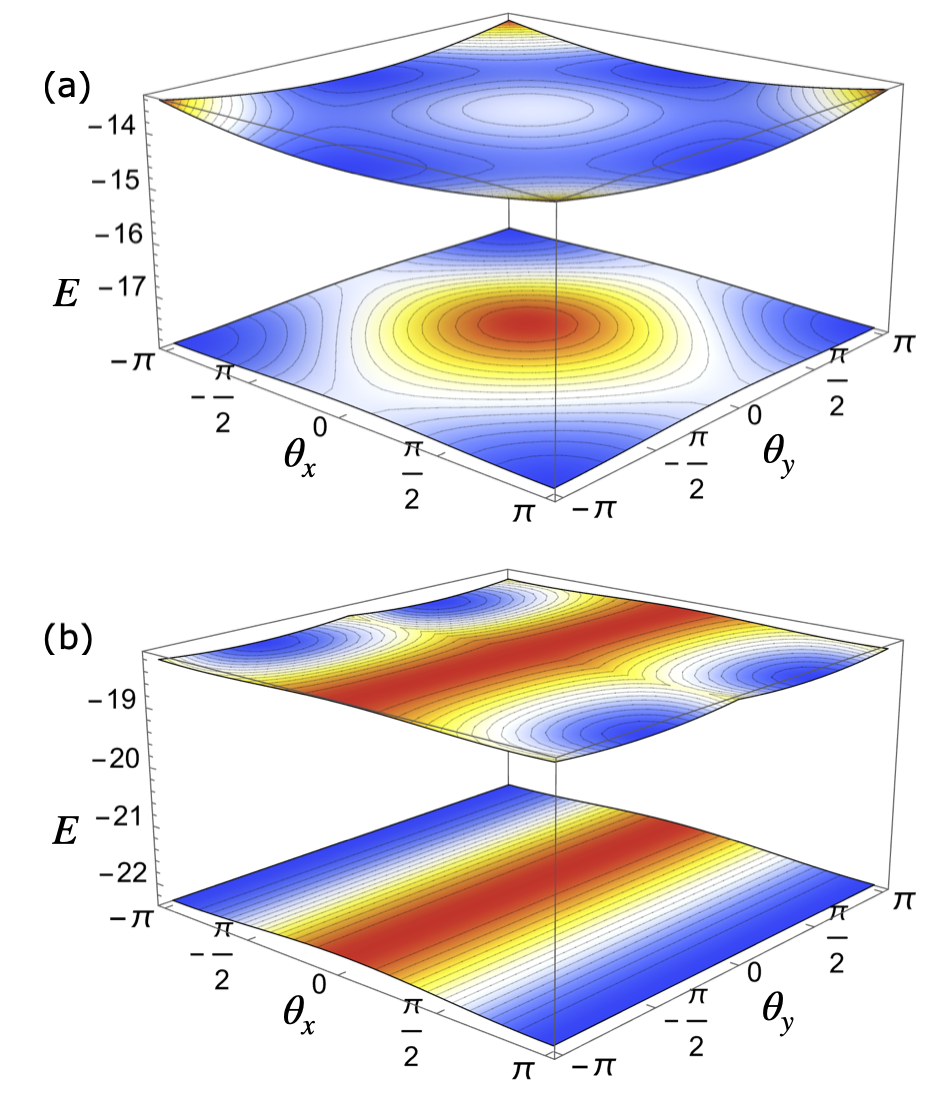}
\caption{$\vec{\theta}$-resolved energy spectrum of $\hat{H}(\vec{\theta})$ with $\varphi = 1/2$ for the two lowest eigenstates for (a) an LSM $4 \times 5$ cluster and (b) a non-LSM $4 \times 4$ cluster. The spectrum of the LSM cluster is periodic in $\theta_y$ with a periodicity of $\pi$. For better clarity, the energies are plotted with a color gradient, where red denotes the global maximum and blue the global minimum.}
 \label{fig:tbc_energy_spectra}
\end{figure}

Next we consider a $4 \times 5$ cluster, which is an LSM cluster as $N_y$ is an odd number. The Hamiltonian now transforms under translation, according to Eq.~(\ref{Eq.many-body_magn_trans_algebra}), as
\begin{equation}
    \hat{T}_{\vec{x}}^{M} \hat{H}(\theta_x, \theta_y) \left( \hat{T}_{\vec{x}}^{M} \right)^{\dagger} = \hat{H}(\theta_x, \theta_y + \pi),
\end{equation}
which is reflected in the $\vec{\theta}$ dependent many-body energy spectrum, see Fig.~\ref{fig:tbc_energy_spectra}. As a result, any single state with an energy gap for all twist angles must have a non-trivial Chern number according to Eq.~(\ref{Eq:universal_relations_many_body}). In addition, the ground state degeneracy of the CDW-phase increases, as a checkerboard pattern is not commensurable with the lattice dimensions anymore. In the limit of $t \to 0$, $V = 1$ there is a tenfold ground state degeneracy with an excitation gap of $V$. This can be understood from the fact that the reciprocal lattice vector $\vec{K} = (\pi, \pi)$ is not in the reciprocal lattice of clusters for odd $N_y$.

\begin{figure}[ht]
    \centering
    \includegraphics[width=1.0\linewidth]{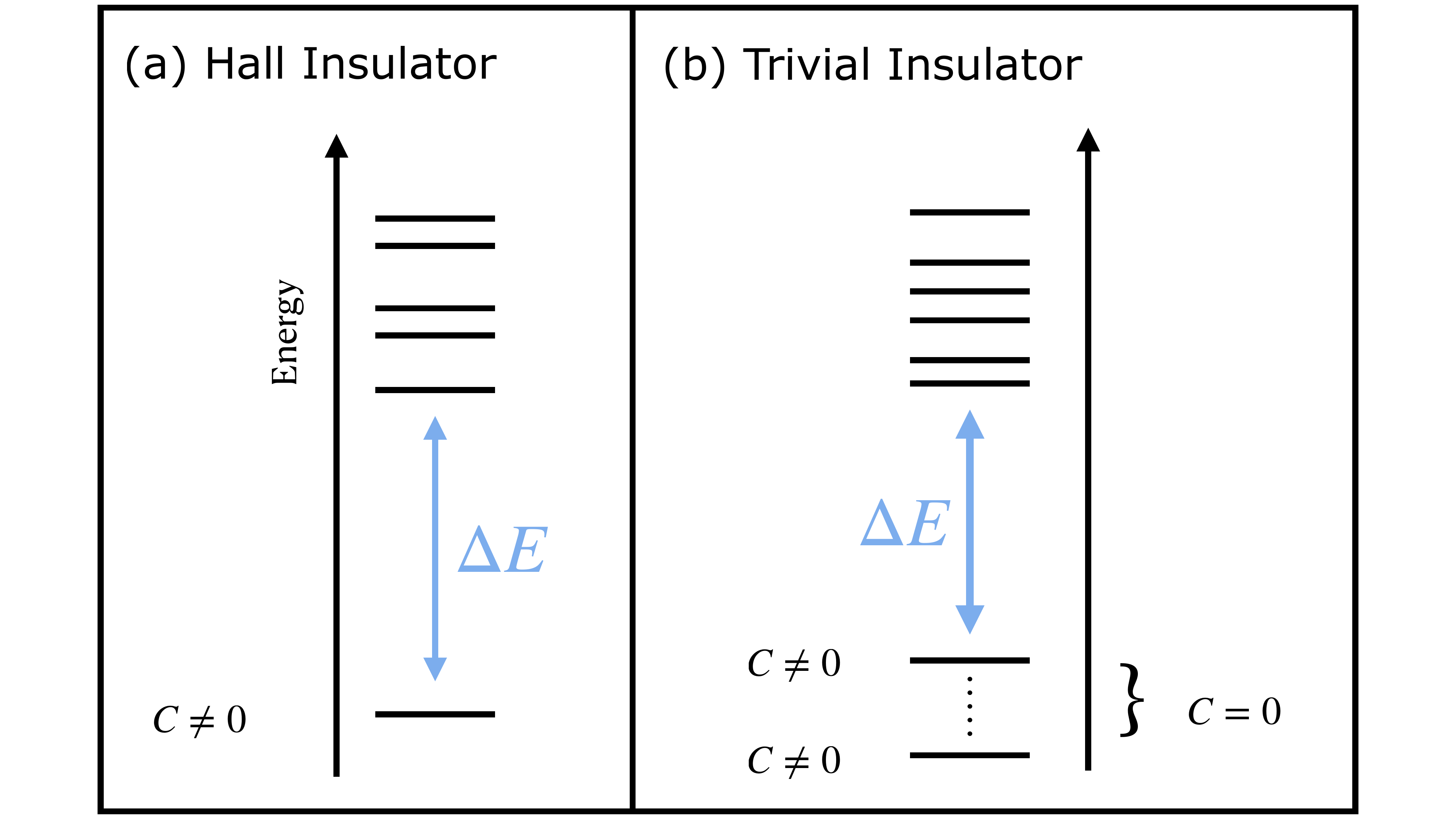}
    \caption{Ground state and ground state quasi-degeneracy of a (a) Hall insulating versus (b) trivial insulating state. In the trivial insulating state, each gapped quasi-degenerate ground state possesses a nontrivial Chern number. In both cases the ground state(s) have to be separated by some energy gap $\Delta E$ from excitations.}
    \label{fig:DiscreteEnergyLevels}
\end{figure}

How does the phase transition occur and can we still speak of a topological phase transition? The system is driven into the CDW by an avoided crossing, where the non-degenerate Hall insulating state turns into ten quasi-degenerate CDW states. The system undergoing an avoided crossing instead of a level crossing is symmetry protected by the LSM-type theorem~\footnote{To be more precise, the LSM-type theorem rules out a topological phase transition to a trivial non-degenerate ground state, but a level crossing protected by rotation symmetry or by translation symmetry $\hat{T}^{M}_{\vec{a}}$ (provided the translation operator commutes with the Hamiltonian) leading to another topologically non-trivial non-degenerate ground state may still be possible.}. For larger LSM clusters, the discontinuous nature of the phase transition that is evident for the non-LSM clusters studied before will become more apparent. While it is true that the lowest energy ground state and the other quasi-degenerate states (assuming that they are all gapped with respect to each other) are individually topologically non-trivial, the sum of their Chern numbers must be equal to zero as they are adiabatically connected to Chern-trivial states at $t \to 0$, $V = 1$. This situation is displayed in Fig.~\ref{fig:DiscreteEnergyLevels}. In this sense the phase transition is still topological, which will be discussed more carefully in Sec.~\ref{sec:fourth_section}.

 \begin{figure}[ht]
 	\centering
 	\includegraphics[width=1.0\linewidth]{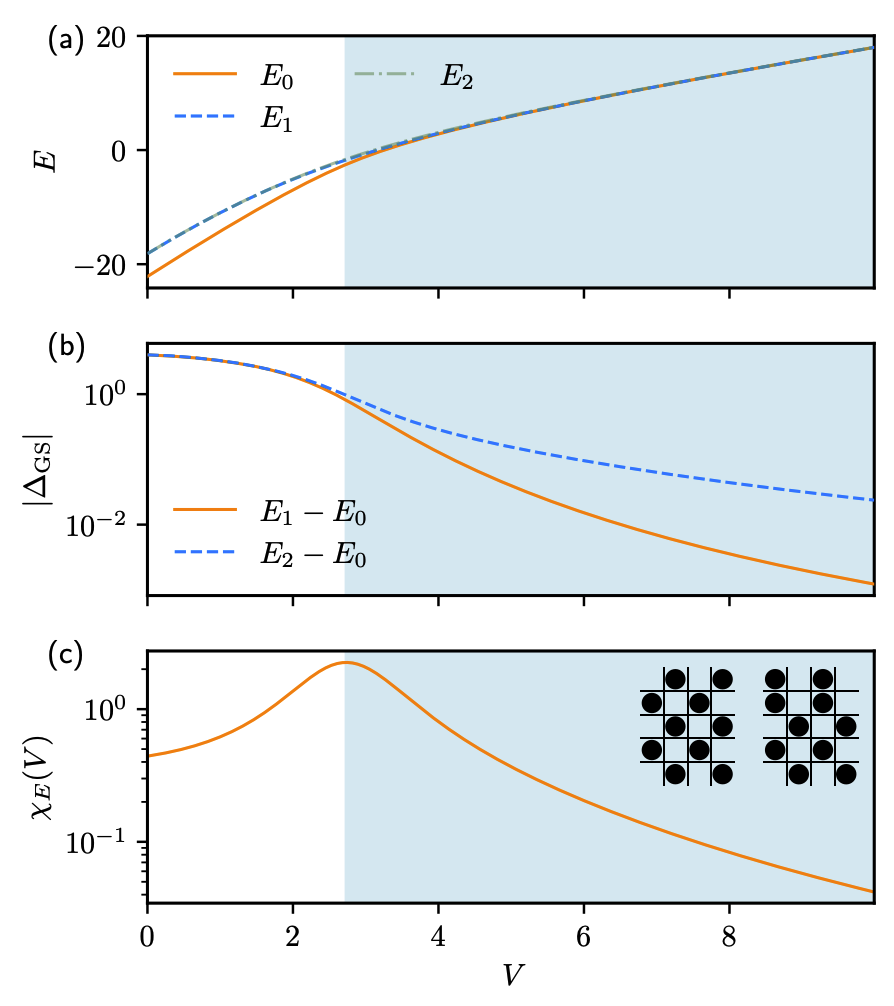}
 	\caption{ED of an LSM $4 \times 5$ cluster. (a) Energy of the ground state $E_0$ and first two excited states ($E_1, E_2$). (b) Energy gap between $E_0$ and ($E_1, E_2$).  (c) Second derivative of $E_0$ as a function of the interaction strength $V/t$. The interaction strength is varied in steps $\delta V = 4/1000$. The IQHE and CDW phases are displayed in white and blue, respectively. The two images correspond to two orthogonal states that minimize $\hat{H}_{\text{int}}$ and that are related to each other by two virtual hoppings.}
 	\label{fig:CDW_Hofstadter_4_5_1_2}
 \end{figure}
 
Calculations are shown in Fig.~\ref{fig:CDW_Hofstadter_4_5_1_2}. Since the LSM-type theorem prohibits a topological phase transition of the lowest energy ground state, there is no obvious choice for a particular twist angle $\vec{\theta}$ anymore. We pick $\vec{\theta} = (0, -\pi  \varphi  N_y)$, where the Hamiltonian is at least $C_2$ invariant. The phase transition is still indicated by a peak of $\chi_E(V)$ at $V = 2.71 \pm 0.01$, although due to strong finite size effects and the cluster not being commensurable to the CDW order, it appears broad on the logarithmic scale. It should be noted that even at $V = 10$ there are still considerable energy gaps between the three quasi-degenerate ground states. This can be understood by considering perturbation theory. The two ground states of a commensurate cluster are related by $N_e$ virtual hoppings with an amplitude of $t/V$, whereas in a non-commensurate cluster hybridization between some of the ground states will approximately take $\sqrt{\rho N_e}$ virtual hoppings, as the CDW order is incommensurable with the lattice geometry by one row, leading to a much larger splitting between the lowest energy levels.

\begin{figure*}[ht]
\subfloat[]{
  \includegraphics[width=0.3275\linewidth]{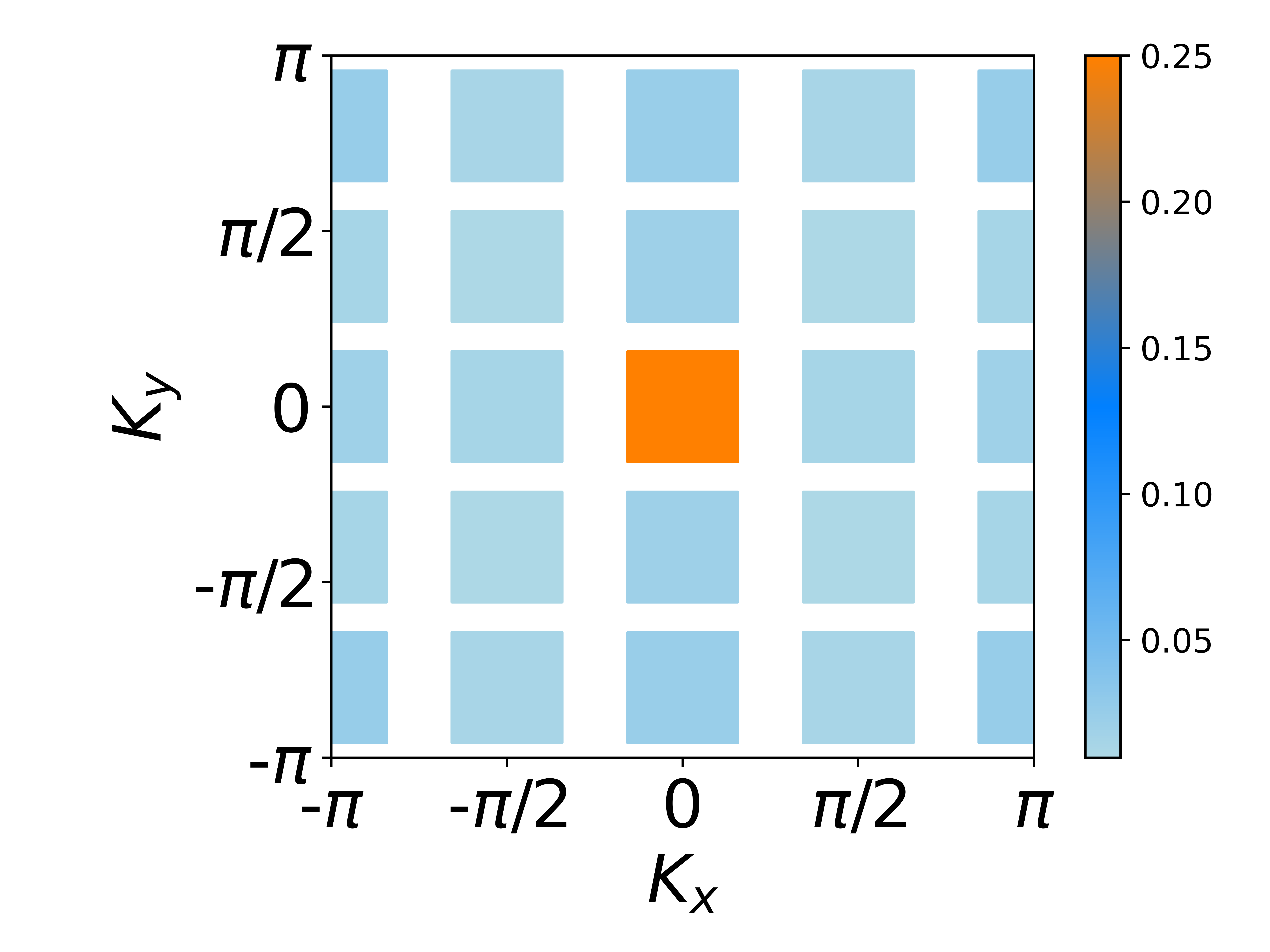}
}
\subfloat[]{
  \includegraphics[width=0.3275\linewidth]{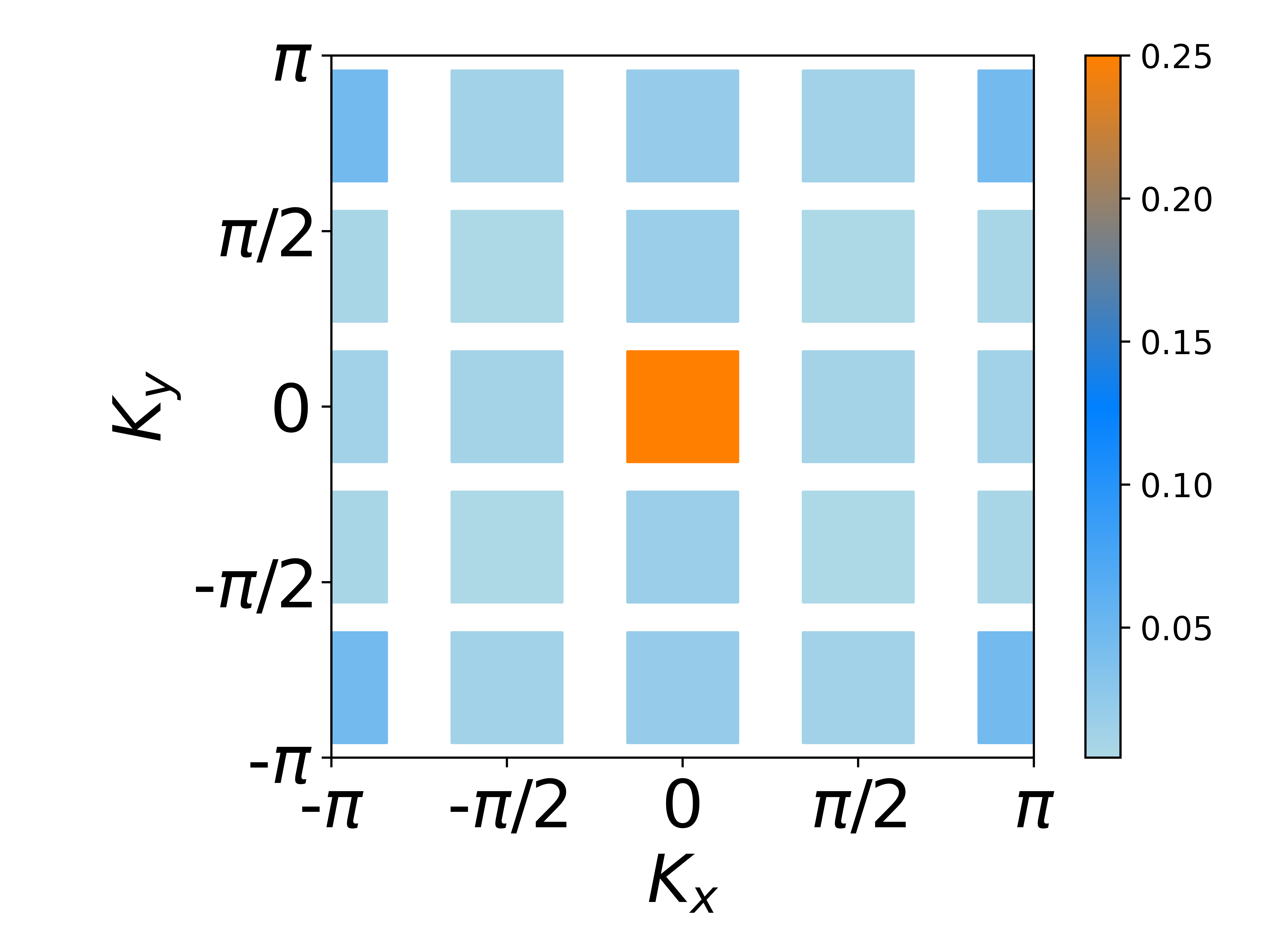}
}
\subfloat[]{
  \includegraphics[width=0.3275\linewidth]{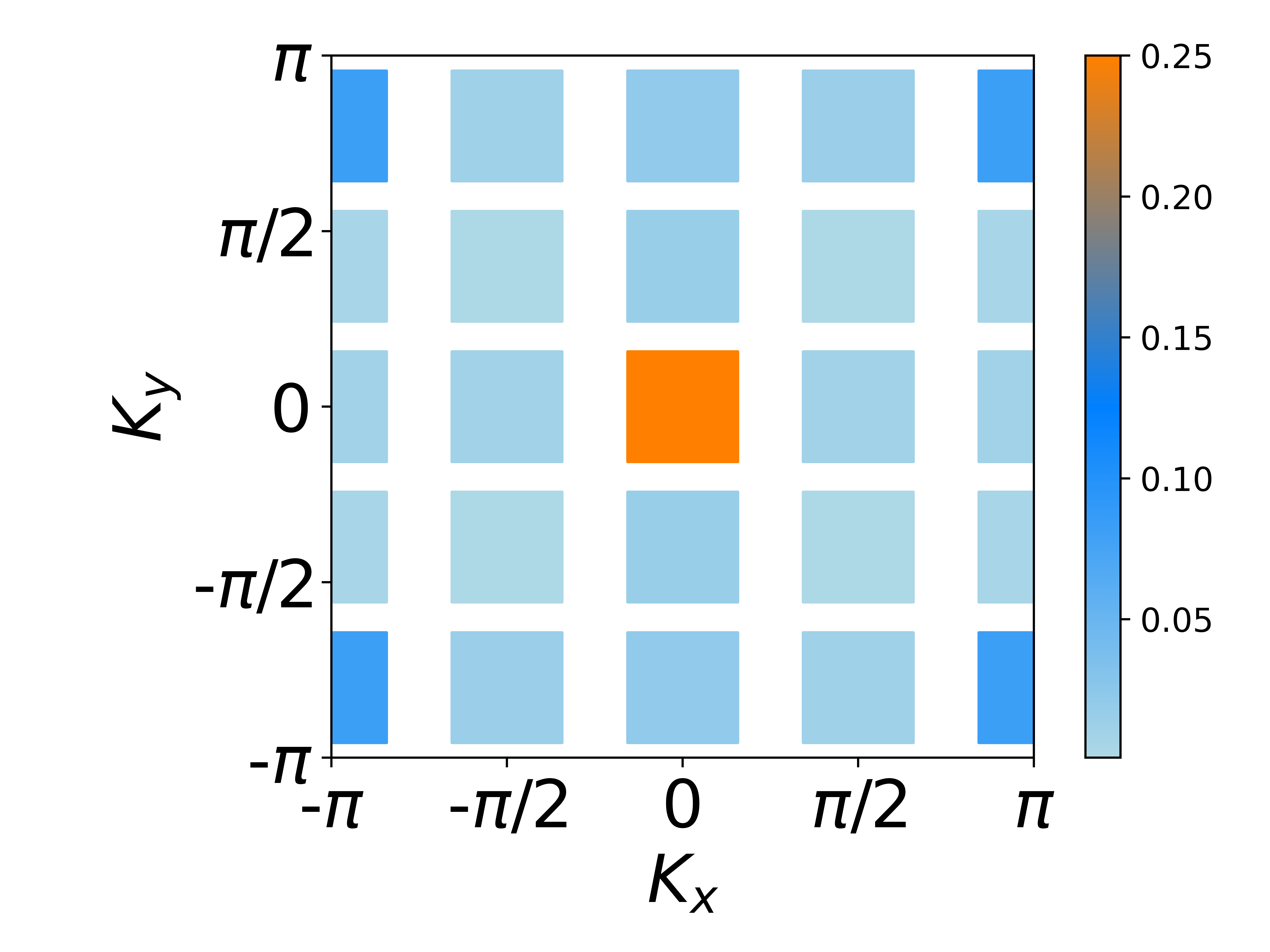}
}

\caption{Static structure factor of the Hofstadter Model on a $4 \times 5$ cluster at (a) $V=2$, (b) $V=4$ and (c) $V=\infty$. Corresponding to the cluster dimensions, the wave vectors of the reciprocal lattice are given by $K_x = - \pi, -\pi/2, \dots, \pi/2$ and $K_y = -4\pi/5, -2\pi/5, \dots, 4\pi/5$.}

\label{fig:Hofstadter_4_5_1_2_SSFs}
\end{figure*}

In Fig.~\ref{fig:Hofstadter_4_5_1_2_SSFs} we display calculations of the CDW structure factor at all $\vec{K}$ points corresponding to the $4 \times 5$ cluster. Since $\vec{K} = (\pi, \pi)$ is no longer contained, the CDW structure factor for $V = \infty$ shows a peak at $K_x = \pi$, $K_y = \pm 4\pi/5$ of $S_{\text{CDW}}(\vec{K}) = 0.082(6)$. While charge order is clearly indicated at large $V$, $S_{\text{CDW}}(\vec{K})$ grows more slowly than in the commensurate non-LSM clusters studied before. Some order does emerge after the phase transition at $V = 4$ indicated by $S_{\text{CDW}}(\vec{K}) = 0.045(9)$ compared to $V = 2$ with $S_{\text{CDW}}(\vec{K}) = 0.025(4)$, however even at $V = 20$ one finds $S_{\text{CDW}}(\vec{K}) = 0.053(6)$, which is still far away from the value at $t = 0$, $V = 1$.

Comparing our results from the non-LSM clusters with the LSM cluster, the discontinuous phase transition into a topologically trivial CDW phase is less clear indicated in the LSM cluster, as the level crossings in the non-LSM clusters lead to a jump of the CDW structure factor and a change of the lowest energy ground state topology. In comparison, the avoided level crossing with continuous $S(\vec{K})$ and $\chi_{E}$ of the LSM cluster makes the identification of phase boundaries more challenging, as well as the fact that the change in the topology of the system can only be described by considering all quasi-degenerate ground states together. In conclusion, the increased ground state degeneracy due to the incommensurability of the checkerboard charge order with LSM clusters and the loss of rotation symmetry as well as the symmetry protected avoided crossing are finite size effects: this implies that the physics of the phase transition in the thermodynamic limit is more adequately described by non-LSM clusters, which clearly indicate a first order transition.

We close this section by appending a comment on the interplay between the topological phase transition of the ground state wave function, the CDW transition and the symmetries of the cluster. Using ED calculations, it is possible to map out the phase boundaries at zero temperature of the Hamiltonian with respect to some parameter using the lowest energy ground state. The topological phase transition may then not occur at the same parameter as the CDW transition transition solely due to finite size effects; in particular this is often caused by the cluster not having the same symmetries as the infinite plane~\cite{Var, Sha}. In our case there will be no topological phase transition of the lowest energy ground state wave function at all for LSM clusters, where the Hamiltonian does not commute with the translation operator, while we found perfect overlap between both transitions for non-LSM clusters. We have already argued that for the characterization of the topological phase transition one has to consider the topology of all quasi-degenerate ground states, instead of the lowest energy one. On non-LSM clusters however, the topological phase transition is also accompanied, by a topological phase transition of the lowest energy ground state by a level crossing. One arrives therefore at the seemingly paradoxical but by no means contradictory result that in the case of actual translation invariance, the Chern number of the ground state is not constrained by the LSM-type theorem, whereas in the other cases the LSM-type theorem is strictly valid and one finds a topological phase---that however is protected only by a finite size gap.

\section{Topological Phase Transitions for Various Flux Quantum Ratios}\label{sec:third_section}

We want to complete the numerical study of Sec.~\ref{sec:second_section} and inquire the possibility of topological phase transitions of the Hofstadter model for flux quantum ratios $\varphi\neq 1/2$ and particle densities $\rho \neq 1/2$ that can still be explored within the limits of ED. In contrast to other methods, ED has the advantage that the aforementioned influence of the cluster geometry on the validity of the LSM-constraint can be inquired; in addition, a mean-field ansatz is difficult to implement for band fillings other than $1/2$, where the order parameter is unknown. This is already an issue for the Hofstadter model on the honeycomb lattice~\cite{Mis}, but especially problematic on the square lattice.

In the following, the Hamiltonian is always of the form
\begin{equation}\label{Eq:general_hofstadter}
    \hat{H} = \hat{H}(\vec{\theta}) + \hat{H}_{\text{int}}(V),
\end{equation}
where $\hat{H}(\vec{\theta})$ has been defined in Eq.~(\ref{Eq:Hofstadter_torus}). NNN-hopping is not included as NN-hopping is sufficient to break TRI. Here $\hat{H}_{\text{int}}(V)$ provides the long-range Coulomb interaction
\begin{equation}\label{Eq:long_range_coulomb}
    \hat{H}_{\text{int}}(V) = \displaystyle\sum_{m \leq m' = 1}^{N_x} \displaystyle\sum_{n \leq n' = 1}^{N_y} \frac{V}{r_{\text{min}}} \hat{n}_{m, n} \hat{n}_{m', n'}
\end{equation}
where $r_{\text{min}}$ is taken to be the smallest distance on the torus geometry. The interaction range is now truly long range and decays like $1/r$, in contrast to Eq.~(\ref{Eq:short_range_coulomb}), where nearest neighbor Coulomb interaction was considered to be sufficient to obtain a CDW at half filling.

We perform calculations for the flux quantum ratios $p/q = 1/5$ and $p/q = 2/5$ with a particle density corresponding to both one or two occupied Hofstadter bands. The critical interaction strengths for each flux quantum ratio and number of particles as well as the single particle gap $\Delta E$ of the non-interacting band structure is displayed in Tab.~\ref{table:critical_U}. As a general trend, smaller single particle bandwidths correspond to smaller $V_{\text{crit}}$. Plots of the single particle band structures are presented in Appendix~\ref{App:single_particle}.

Plots for the static structure factors are provided in Appendix~\ref{App:static_structure} for the closest numerically evaluated interaction strengths before and after the phase transition occurs into the CDW regime. The onset of charge order is clearly indicated by the jump of $S_{\text{CDW}}$ at several $\vec{K}$ values. In addition, for clusters with $\rho = 2/5$ we observe a breaking of $C_4$ rotation symmetry.


\begin{table}[h!]
\centering
\begin{tabular}{|c | c | c | c | c |} 
 \hline
 $p/q$ & $N_e$ & $N_x \times N_y$ & $\Delta E$ &  $V_{\text{crit}}$ \\ [0.5ex] 
 \hline\hline
 1/5 & 5 & $5\times 5$ & 1.553699 & $22.652 \pm 0.002$\\
 1/5 & 10 & $5\times 5$ & 0.520147 & $4.7875 \pm 0.0125$\\
 2/5 & 5 & $5\times 5$ & 0.157179 & $4.918 \pm 0.002$\\
 2/5 & 10 & $5\times 5$ & 1.539145 & $17.775 \pm 0.025$\\
 [1ex] 
 \hline
\end{tabular}
\caption{Critical interaction strengths $V_{\text{crit}}$ and single-particle band gaps $\Delta E$ for varying flux quanta, particle numbers and lattice dimensions. The particle number $N_e$ always corresponds to an integer filling of non-interacting bands.}
\label{table:critical_U}
\end{table}

\section{Spontaneous translation invariance breaking}\label{sec:fourth_section}

In view of the previous numerical results, in this section we want to inquire the relation between spontaneous translation symmetry breaking and the topological phase transition from Chern to trivial insulators with magnetic translation symmetry more rigorously, aware that the LSM-type theorem is a non-perturbative result. It is clear that for macroscopically large systems, the physics behind the phase transition cannot change in a meaningful way depending on the specific cluster geometry. Nevertheless, in the following argumentation we will have to carefully distinguish between LSM- and non-LSM clusters, by virtue of the fact that some properties are only mathematically well-defined in one or the other. An example is the LSM-type theorem itself, which strictly holds only for LSM-geometries. Nonetheless, the physical relevance of these properties has to be the same in the thermodynamic limit, independent of the lattice geometry.

We start by considering integer fillings---that is an integer number of Hofstadter bands are occupied. These cases have been investigated numerically in this work. Then we continue with the consideration of fractional fillings and the differences between lattice and continuum models in this context. For convenience, when translation symmetry is addressed, we always have the \textit{magnetic} translation operator $\hat{T}_{\vec{a}}^{M}$ in mind unless stated otherwise.

\subsection{Integer quantum Hall effect}\label{subsec:first_subsec}

The formation of a CDW phase for a translation invariant Hamiltonian requires spontaneous symmetry breaking of translation invariance. In the thermodynamic limit, the Hamiltonian will obtain a set of degenerate ground states. For example, in Sec.~\ref{sec:second_section} we have studied the formation of two quasi-degenerate states that display a checkerboard pattern. The wave function that describes a real sample (at zero temperature) will be a superposition of these ground states that breaks translation symmetry, in the case of the two checkerboard states by being localized on the ``black'' or ``white'' lattice sites. In compliance with our numerical calculations we want to understand this mechanism and its physical implications on the LSM-type theorem now for possibly macroscopic systems, but still of finite size. We need to understand why quasi-degeneracy is essential for the phase transition into a topologically trivial insulating state and why its origin must be given by the spontaneous breaking of translation symmetry as opposed to some other cause.

On non-LSM clusters, i.e. clusters where the magnetic translation operators commute with the Hamiltonian, spontaneous breaking of translation symmetry must be accompanied by a set of quasi-degenerate ground states that are eigenstates of the translation operators with different crystalline momenta $\vec{K}$. While the different merely quasi-degenerate eigenstates (in particular the ground states) of the Hamiltonian are naturally eigenstates of the translation operator for these finite size clusters, a superposition of states belonging to different $\vec{K}$ breaks translation invariance. 

\noindent $\bullet$ {\it Symmetry breaking and local order parameter.} Why does the ground state wave function of a real sample break translation symmetry? In the following we will assume that the symmetry breaking phase also has a local order parameter, such as the varying charge density. Then the macroscopic wave function will collapse upon any measurement of the local order parameter into one of the symmetry broken states. The system is trapped in the symmetry broken state~\footnote{More precisely, ergodicity is broken as the tunneling amplitude from a symmetry broken state to another one due to adiabatic flux insertion vanishes exponentially fast for a lattice Hamiltonian. In the continuum collective excitations may actually drive the system into a different state.}. A flux insertion to measure the Hall conductivity within the linear response regime is too small a perturbation to let the system tunnel into another ground state, in particular if the broken symmetry is discrete. Note that this argument relies on the existence of a local order parameter, a point to which we will come back to in the next subsection.

We have seen that the CDW phase comes with a quasi-degenerate ground state spectrum and a translation symmetry breaking state best describes the real system. How is this connected to topology? On non-LSM clusters Eq.~(\ref{Eq:universal_relations_many_body}) does not strictly hold anymore. This manifests itself in the fact that in that case it is possible to have gapped eigenstates of $\hat{H}(\vec{\theta})$ for all twist angles $\vec{\theta}$ with a trivial many-body Chern number, which justified the use of symmetry-indicators to detect the topological phase transition in Sec.~\ref{sec:second_section}.

\noindent $\bullet$ {\it Necessity of quasi-degeneracy.} Is a ground state quasi-degeneracy then required to set up a topologically trivial phase and does it have to be related to spontaneous symmetry breaking of translation symmetry? Let us first address the question of degeneracy. In the case of LSM clusters they must have a quasi-degeneracy of a multiple of $q$ according to Eq.~(\ref{Eq:universal_relations_many_body}), if they are supposed to belong to a trivial insulating phase, otherwise the sum of the Chern numbers of the ground states could not be zero and the ground states would not be adiabatically connected to localized, hence topologically trivial CDW states. On non-LSM clusters that have no long-range structure of period $q$~\footnote{In contrast, a long-range structure of period $q$ could be provided for example by a stripe pattern of impurities that would only break translation symmetry in one direction and would in fact result in a non-degenerate and topologically trivial ground state if the stripe pattern is commensurable with the given cluster.} one also expects to find quasi-degeneracy~\cite{Osh}. If the physical origin of the quasi-degeneracy is the formation of a CDW, then the number of ground states may in fact only change depending on whether the CDW pattern is commensurable with the given cluster as we have seen in Sec.~\ref{sec:second_section}. However, the very presence of many ground states must pertain to all possible impurity-free clusters.

\noindent $\bullet$ {\it Degree of quasi-degeneracy.} Why is the degeneracy a multiple of $q$ for non-LSM clusters? Consider mean-field theory: if the formation of a CDW is modeled by an on-site potential of an effective single-particle Hamiltonian, then the unit cell of the effective Hamiltonian must contain an integer multiple of a flux quantum for the single particle band structure to be topologically trivial according to Eq.~(\ref{Eq:universal_relation}). This is only possible if it contains a multiple of $q$ original unit cells plaquettes resulting in an integer multiple of $q$ different possible positions of the on-site potential and hence an integer multiple of $q$ different ground states. Note that one considers different mean-field Hamiltonians for a single Hamiltonian of the many-body system. The ground states are degenerate, as the different mean-field Hamiltonians are equivalent to each other under translation on non-LSM clusters.

\noindent $\bullet$ {\it Uniqueness of translation symmetry breaking.} Why is the spontaneous breaking of the translation symmetry the origin of the quasi-degeneracy in a topologically trivial phase?
We have argued, why a quasi-degeneracy of ground states in a translation invariant Hamiltonian in an external magnetic field is essential to have a trivial Hall conductivity. An example for a different discrete symmetry that may be broken spontaneously is TRI, which can in principle occur at a flux quantum ratio of $p/q = 1/2$, where the Hamiltonian commutes with the TR operation if one only includes nearest neighbor hoppings in the Hamiltonian. In case of spontaneous TRI breaking, the Hamiltonian would have two time reversal partners as ground states that do not have to break translation symmetry. In Ref.~\cite{Lu}, Eq.~(\ref{Eq:universal_relations_many_body}) is argued to hold in the thermodynamic limit if the ground state is an eigenstate of the magnetic translation operator. If translation maps a flux (associated to some twist angle) to a different flux according to Eq.~(\ref{Eq.many-body_magn_trans_algebra}), then ground states possess an effective translation symmetry, expressed through the modified translation operator
\begin{equation}\label{Eq:effective_symmetry}
    \Tilde{\hat{T}}^{M}_{\vec{x}} = \hat{F}_y(2\pi \varphi N_y) \hat{T}^{M}_{\vec{x}},
\end{equation}
where $\hat{F}_y(2\pi \varphi N_y)$ is a flux insertion to exactly compensate the change of Aharonov-Bohm phases around non-contractable loops of the given cluster. Let us pick one of the two time reversal partners. Under the tacit assumption that this state cannot transform into its partner by translation or adiabatic flux insertion, the Hall-conductivity would then still be quantized to non-zero values, as the effective translation symmetry maps the state onto itself. A topologically trivial phase hence requires SSB of translation symmetry.

The above argument should hold for systems with a local, gauge invariant order parameter, too. What happens in topologically ordered systems though, where the assumption that a state maps to itself under adiabatic flux insertion is violated? Such a system could be realized in the form of reentrant superconductivity with presumed triplett pairing which has been observed in UTe${}_{2}$~\cite{Ran} and moir\'{e} systems~\cite{Cao}. A BCS superconductor in 2D has a topologically four-fold quasi-degenerate ground state due to the presence of cooper pairs~\cite{Han}; higher ground state degeneracy could be obtained by different pairing mechanisms, such as charge-4e pairing~\cite{Jia2} or, hypothetically, by anyon pairing superonductivity~\cite{Fra}. In the presence of a magnetic field however, these systems will break translation symmetry as a consequence of the breaking of U(1) symmetry~\cite{Sha2}: Since in a two-particle Hilbert space the operation
\begin{equation}
    \hat{T}^{M}_{\vec{x}} \hat{T}^{M}_{\vec{y}} \left( \hat{T}^{M}_{\vec{x}}\right)^{\dagger} \left( \hat{T}^{M}_{\vec{y}} \right) = \e^{4\pi \ci \varphi}
\end{equation}
effectively acts as a U(1) symmetry on the gap function (which transforms like a pair of two particles), magnetic translation symmetry has to be broken in the superconducting state. An exception is provided for $q=2$: Since in the two particle Hilbert space translation operators commute, it is possible to obtain a translation invariant gap function. Filling half a Hofstadter band at $q=2$ would require according to Eq.~(\ref{Eq:fractional_relation}) a fourfold degenerate ground state for a trivial Hall-conductivity, which would be provided by the BCS state. In additian, as argued above, at $q=2$ the Hofstadter Hamiltonian for nearest neighbor hopping has TRI, further suggesting a trivial Hall insulating state. It would be an interesting question for future research, whether such a state can be realized or not.

\noindent $\bullet$ {\it Strong-coupling limit} We want to provide a final proof, why the Hall conductivity for sufficiently strong long-range interaction has to be trivial, independent of the filling factor, lattice geometry and underlying tight-binding Hamiltonian. In the limit of infinitely strong interaction strength, ground states must minimize $\braket{\hat{H}_{\text{int}}}$. The set of ground states will then only consist of charge-ordered states $\ket{\tilde{\Psi}_{i}}$ that do not hybridize with respect to the tight-binding Hamiltonian, i.e. $\braket{\tilde{\Psi}_{i}|\hat{H}|\tilde{\Psi}_{j}} = 0$, as going from one state to the other would require multiple fermions to hop to a different lattice site, provided the interaction is long ranged. As a consequence, ground states will not be affected by a flux insertion so that only zero Hall transport is possible. Finite $t/V \ll 1$ leads to negligible corrections; in the case of $\varphi = 1/2 = \rho$ and non-LSM clusters, for example, these are of the order of $\mathcal{O}(t/V)^{N_e}$.

\subsection{Fractional quantum Hall effect}\label{subsec:second_subsec}

We now discuss the more general case of fractional fillings, in particular the fractional quantum Hall effect (FQHE) of which the Hofstadter model provides an excellent platform for~\cite{Lon, Haf2, Ger, And}. Here, a genuine many-body treatment is inevitable as the presence of quasiparticles in a topologically ordered system does not permit the usual mean-field ansatz~\footnote{This does not rule out the possibility of a mean-field ansatz in terms of an effective quasiparticle Hamiltonian~\cite{Zha}}. In contrast to integer fillings, ground state degeneracy is enforced by the generalized LSM theorem in the fractional Hall insulating phase and the LSM-type theorem becomes~\cite{Mat}
\begin{equation}\label{Eq:fractional_relation}
    \e^{2\pi \ci (\frac{p}{q} C - \rho) D} = 1,
\end{equation}
with $q \rho = \bar{m}/D$ and coprime $\bar{m}, D$, where $D$ is equal to the ground state degeneracy and $C$, which is now to be interpreted as the Chern number per ground state, is a rational fraction of $D$. We will argue why the characteristic topological properties of these ground states, among others the LSM-type theorem Eq.~(\ref{Eq:fractional_relation}), are protected by magnetic translation symmetry before arguing why CDW states are trivial nevertheless.

One of the hallmarks of the FQHE is topological order that manifests itself on a torus geometry as ground state degeneracy due to the presence of quasiparticles in the thermodynamic limit~\cite{Osh2}. The number of ground states is independent of the chosen cluster as its origin lies in the quasiparticle statistics and not in some long-range structure in real space. For the calculation of the Hall conductivity, the joint Chern number of all ground states has to be computed, in contrast to the integer filling states in Sec.~\ref{subsec:first_subsec}, where we argued that a flux insertion should be seen as a small perturbation that does not allow a ground state to tunnel into a different ground state. However, in the case of the FQHE, the different ground states have to transform into each other under flux insertion as argued in~\cite{Wen2, Osh2} from a field-theoretic perspective. Let
\begin{equation}\label{Eq:flux_insertion_operators}
    \tilde{\hat{F}}_{x/y}(\Phi_0)
\end{equation}
be a combination of a flux insertion operation of $\Phi_0$ and a gauge transformation (see Eq.~(\ref{Eq:combined_flux_insertion_gauge_trafo})) that maps a ground state to another ground state. Then
\begin{equation}\label{Eq:flux_operators_algebra}
    \tilde{\hat{F}}_x(\Phi_0) \tilde{\hat{F}}_y(\Phi_0) \cong \e^{-2\pi \ci /D} \tilde{\hat{F}}_y(\Phi_0)  \tilde{\hat{F}}_x(\Phi_0).
\end{equation}
The symbol $\cong$ restricts Eq.~(\ref{Eq:flux_operators_algebra}) to the ground state subspace. Eq.~(\ref{Eq:flux_operators_algebra}) implies at least $D$ ground states as the flux insertion operators obey the same algebra as magnetic translation operators. This transformation behavior as well as the ground state degeneracy~\cite{Hal} can be proven exactly for some non-LSM clusters as demonstrated in Appendix~\ref{App:first_appendix}, even though, similarly to the generalized LSM theorem and LSM-type theorem, there are some technicalities depending on the chosen cluster. In particular, as any stable exact degeneracy has to be symmetry protected, Eq.~(\ref{Eq:flux_operators_algebra}) strictly holds only if the topologically degenerate ground states (that is degenerate in the thermodynamic limit even in the presence of impurities or lattice potential~\cite{Wen2}) are related by magnetic translation symmetry (see Eq.~(\ref{Eq:fractional_translation}) and Eq.~(\ref{Eq:fractional_translation_phase_factor})). Otherwise finite size splittings lead to a change of the algebra in Eq.~(\ref{Eq:flux_operators_algebra}), provided the adiabatic flux insertion is slow compared to finite size gaps, which is not an issue in the thermodynamic limit. Since the different states transform into each other, when inserting flux to measure the Hall conductivity, the contribution of all ground states has to be summed.

What happens in the CDW phase regime? In the limit $t \to 0$, $V = 1$ it is clear that CDW states don't transform under a flux insertion and should hence be eigenstates of both flux insertion operators, implying that the flux insertion operators in Eq.~(\ref{Eq:flux_insertion_operators}) now commute. This already implies at least $D^2$ quasi-degenerate ground states, compared to the $D$ ground states of the FQHE, see Appendix~\ref{App:first_appendix}. In addition, the number of ground states also has to be an integer multiple of $q$, according to Eq.~(\ref{Eq:fractional_relation}). The quasi-degeneracy turns into sets of $D$ exactly degenerate ground states if they are related by translation, as then they transform under irreducible projective representations $D_{M}^{\vec{K}}$ of the translation group~\cite{Bro}. It should be noted that these degenerate states possess different crystalline momenta, when expressed in the eigenbasis of either $\hat{T}_{\vec{x}}^{M}$ or $\hat{T}_{\vec{y}}^{M}$. This alone, however, does not indicate spontaneous symmetry breaking. In the FQHE regime this difference in momentum is carried by quasiparticles known as visons in the context of quantum Hall physics~\cite{Lu, Par}. Instead, in the CDW phase regime there have to be $D$ copies of each $D_{M}^{\vec{K}}$. In the case that the many-body translation operators commute and states transform under the usual one dimensional irreducible representations of the translation group, the symmetry protection of the ground state degeneracy is lifted and only maintained in the thermodynamic limit. The ground states of the FQHE then all carry the same momentum, whereas the CDW are again characterized by different momenta~\cite{Par}.

\subsection{Continuous space}\label{subsec:third_subsec}

So far only lattice Hamiltonians have been considered. We close this section by touching briefly on free interacting particles. The absence of a periodic lattice potential implies continuous translation invariance in the infinite plane. On a torus, following the argumentation leading to Eq.~(\ref{Eq.many-body_magn_trans_algebra}), a twist angle $\vec{\theta}$ can be transformed into any other twist angle by a suitable translation. This property is independent of the chosen sample dimensions so that the LSM-type theorem always applies exactly. In fact, translation invariance implies that the Chern numbers of each Landau level have to be equal to one, which can be seen in the following way. It is possible to divide a torus, being pierced by $N_{\Phi_0}$ unit flux quanta, into a grid of $N_c$ virtual plaquettes with a flux of $p/q$ flux quanta per plaquette, so that $N_{\Phi_0} = N_c p/q$. A filling factor of $n_{LL}$ occupied Landau levels leads to an average particle density of
\begin{equation}
    \rho = \frac{N_{\Phi_0} n_{LL}}{N_c} = \frac{p}{q} n_{LL}.
\end{equation}
particles per virtual plaquette With coprime $p,q$ and according to Eq.~(\ref{Eq:universal_relations_many_body}) this immediately implies that
\begin{equation}\label{Eq:chern_ll}
    C - n_{LL} = 0 \text{ mod } q.
\end{equation}
Since this argument is independent of the partition of the torus, Eq.~(\ref{Eq:chern_ll}) has to hold for all $q$, leading to $C = n_{LL}$.

At sufficiently large long-range interactions, a quantum Hall system can be expected to be driven into a Wigner crystal phase with zero Hall conductance~\cite{Tes}. In case the classical Wigner lattice does not contain an integer number of flux quanta, even in the limit of strong coulomb interaction, finite kinetic energy may cause the electrons to condense into stripe or bubble patterns that may lead to further reduction of translation symmetry~\cite{Ct, Fog}. From the previous paragraph it follows that either the number of ground states would have to be infinite which can be ruled out for finite clusters or there has to be a gapless excitation spectrum. This is realized, at least in the low energy sector, by Goldstone modes as a consequence of the breakdown of continuous translation symmetry: in a Wigner crystal electrons order periodically and hence possess gapless phonon-like excitations~\cite{Ban}. The stability of such a charge ordering in the case of free particles depends on the electrons being subject to a long-range interaction (especially at finite temperature, where long range interaction provides a loophole of the Hohenberg-Mermin-Wagner theorem~\cite{Halp}). Indeed, in the case of the FQHE sufficiently short ranged interaction actually stabilizes the incompressible quantum Hall liquid state~\cite{Tru, Haf2}.

It should also be noted that the breakdown of translation invariance is a necessary but no sufficient criterium for a trivial Hall conducting phase. Although Wigner crystallization, where the crystallized electrons possess a trivial Hall conductivity, are used to explain transitions from FQH phases to reentrant IQH phases~\cite{Shi, Liu}, nonzero Hall conductivity may coexist with electrons forming crystalline order~\cite{Tes} and can be caused by collective excitations, which is backed by experimental data~\cite{Nar, Yi, Zhu}. The existence of crystalline order itself in quantum Hall systems is by now well established and has been observed directly using high-resolution scanning tunnelling microscopy~\cite{Tsu}. Direct measurements of the Hall conductivity however are impeded by the presence of finite longitudinal resistivity characteristic to Wigner crystal phases. Experimental evidence indicates that Wigner crystals are trivially insulating if they are pinned by impurities~\cite{Mad}. As we have demonstrated in Sec.~\ref{subsec:second_subsec}, a periodic lattice potential can play the same role, which comes to fruition in moir\'{e} systems~\cite{Pad}.

\section{Conclusion and Outlook}\label{sec:Conclusion}

In this paper we have studied topological phase transitions of the interacting Hofstadter model from a Chern insulating to a trivially insulating charge-density-wave phase for various flux quantum ratios and particle numbers. The constraint on the quantization of the Hall conductivity due to Lieb-Schultz-Mattis (LSM)-type theorem, depending on translation symmetry that in particular rule out a Chern-trivial phase, is circumvented via spontaneous breaking of translation symmetry driven by long-range interaction. Qualitative differences between the inquired cluster geometries emerge depending on whether they have to strictly obey the LSM-type theorem. On a large class of clusters the topological phase transition occurs via a symmetry protected level crossing with a jump of the charge density wave structure factor indicating a discontinuous phase transition into a topologically trivial CDW phase. Only on a specific set of cluster geometries it is seen that a topologically non-trivial phase is protected by a finite size gap which however quickly goes to zero in the thermodynamic limit. 

For future studies, it would be interesting to inquire interacting quantum spin Hall phases subject to half integer flux quanta, where LSM-type theorems also impose a constraint on the $Z_2$ invariant~\cite{Wu}, in particular it is far less clear, whether a finite size gap can stabilize a genuine topological many-body phase, as the definition of the $Z_2$ invariant often relies on field-theoretic arguments or a Schmidt decomposition which can be problematic in the case of quasi-degeneracy. A tool that may be more suited to study SSB for larger systems, even though the subtleties of the lattice geometry may be lost, is DMRG, where the obtained ground state usually spontaneously breaks symmetries, if the true ground state is quasi-degenerate~\cite{Jia}. Finally, while the present study focuses on the Hofstadter Hamiltonian as a popular toy model, in particular in the study of cold-atom gases~\cite{Aid}, our results may also be of value for quantum Hall phases in Moir\'{e} superlattices, where the flux quantum per unit cell can be of the order of a unit flux quantum~\cite{Dea, San}.

\begin{acknowledgments}

We wish to thank Elias Lettl and Alexander Wietek for stimulating discussions. This research was supported by the Deutsche Forschungsgemeinschaft (DFG) through QUAST-FOR5249 - 449872909 (project TP4) and it was supported in part by grant NSF PHY-2309135 to the Kavli Institute for Theoretical Physics (KITP).

\end{acknowledgments}

\appendix

\section{Proof of the Hall conductivity constraint}\label{App:zero_appendix}

Here we outline how to deduce Eq.~(\ref{Eq:universal_relation}) and its many-body generalization. The Chern number of Bloch states $\ket{\Psi^{n}(\vec{k})}$ occupying a gapped band is given by
\begin{equation}\label{Eq:Chern_number}
    C_n = -\frac{1}{2\pi} \int_{\text{BZ}} \diff \vec{S} \vec{\nabla}_{\vec{k}} \times \vec{A}^{n}(\vec{k})
\end{equation}
with the Berry connection
\begin{equation}
    \vec{A}^{n}(\vec{k}) = \ci \braket{\Psi^{n}(\vec{k})|\vec{\nabla}_{\vec{k}}|\Psi^{n}(\vec{k})}.
\end{equation}
It is possible to rewrite Eq.~(\ref{Eq:Chern_number}) using Stokes' theorem as a line integral over the boundary of the Brillouin zone~\cite{Fue}. Then the wave function must be chosen to be smooth over the entire Brillouin zone but cannot satisfy the boundary condition of the Brillouin zone, if it has a non-zero Chern number. For the Bloch states defined in the leadup to Eq.~(\ref{Eq:magn_transl_degeneracy}) we can define a smooth gauge with the boundary conditions
\begin{equation}
    \begin{gathered}[b]
        \ket{k_x + 2\pi/q, k_y} =  \ket{k_x, k_y} \\
        \ket{k_x, k_y + 2\pi} =  \e^{-\ci q C k_x} \ket{k_x, k_y}.
    \end{gathered}
\end{equation}
Furthermore, we specify the wave functions to satisfy~\cite{Dan}
\begin{equation}
    \hat{T}_{\vec{x}}^{M} \ket{k_x, k_y} = \e^{-\ci t k_x q} \ket{k_x, k_y - 2\pi \varphi},
\end{equation}
where the integer $t$ will be model specific (in the case of the Hofstadter model with only nearest neighbor hopping, they are given by a specific Diophantine equation~\cite{Fue} as a special case of Eq.~(\ref{Eq:universal_relation})). We then identify
\begin{equation}
\begin{aligned}[b]
    \e^{-\ci q k_x} \ket{k_x, k_y} = \left( T_{\vec{x}}^{M} \right)^{q} \ket{k_x, k_y} = \e^{-\ci t k_x q^2} \ket{k_x, k_y - 2\pi p} \\
    = \e^{-\ci t k_x q^2 - \ci C p q k_x} \ket{k_x, k_y},
\end{aligned}
\end{equation}
leading to
\begin{equation}
    C \varphi - \frac{1}{q} = 0 \text{ mod 1}.
\end{equation}
If we don't occupy a single band, but many bands with a number of particles per unit cell $\rho$, then Eq.~(\ref{Eq:universal_relation}) follows.

For many-body systems, Eq.~(\ref{Eq:universal_relations_many_body}) can be shown analogously~\cite{Mat}. The only major difference is that the many-body translation operators technically depend on twist angles resulting in some additional gauge transformations in the proof. For example, 
\begin{equation}
\begin{aligned}[b]
    \hat{T}_{\vec{x}}^{M, \theta_x} c_{m,n}^{\dagger} \left(  \hat{T}_{\vec{x}}^{M, \theta_x}\right)^{\dagger} = & \e^{-\ci \theta_x \delta_{m, N_x}} \e^{-2\pi \ci \varphi n} c_{m+1,n}^{\dagger}\\
    \hat{T}_{\vec{y}}^{M, \theta_y} c_{m,n}^{\dagger} \left(  \hat{T}_{\vec{y}}^{M, \theta_y}\right)^{\dagger} = & \e^{-\ci \theta_y \delta_{n, N_y}} c_{m,n+1}^{\dagger}
\end{aligned}
\end{equation}
commute with the Hofstadter Hamiltonian in Eq.~(\ref{Eq:Hofstadter_torus}), provided $N_{x/y}$ are integer multiples of $q$. In addition, while for single particles on the infinite plane one can pick any
\begin{equation}
    \left( \hat{T}_{\vec{x}}^{M} \right)^{\alpha}, \left( \hat{T}_{\vec{y}}^{M} \right)^{\beta}
\end{equation}
with $\alpha \beta = q$ to define a set of Bloch vectors (where for convenience one choses $\alpha = q$, $\beta = 1$), in the many-body case the translations that map one twist angle to another one are determined by the cluster geometry. If translations in the $\vec{x}$-direction result according to Eq.~(\ref{Eq.many-body_magn_trans_algebra}) in $\alpha'$ different twist angles and translations in the $\vec{y}$-direction result in $\beta'$ different twist angles, then Eq.~(\ref{Eq:universal_relations_many_body}) follows if $\alpha' \beta' = q$. Otherwise for $\alpha' \beta' < q$ one obtains a weaker quantization condition or none in the case of $\alpha' = \beta' = 1$.

\section{Single-particle band structures}\label{App:single_particle}

\begin{figure*}
\subfloat[]{
  \includegraphics[width=0.45\linewidth]{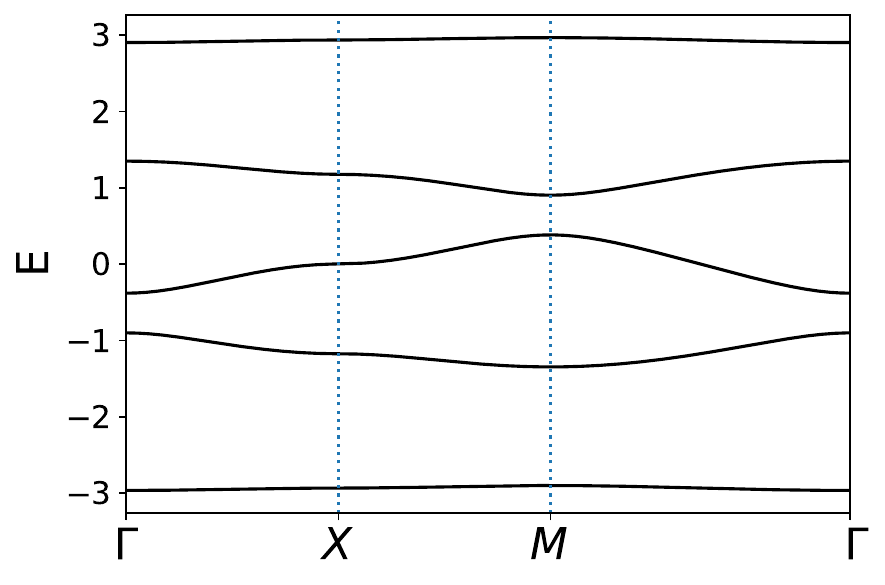}
}\hfil
\subfloat[]{
  \includegraphics[width=0.45\linewidth]{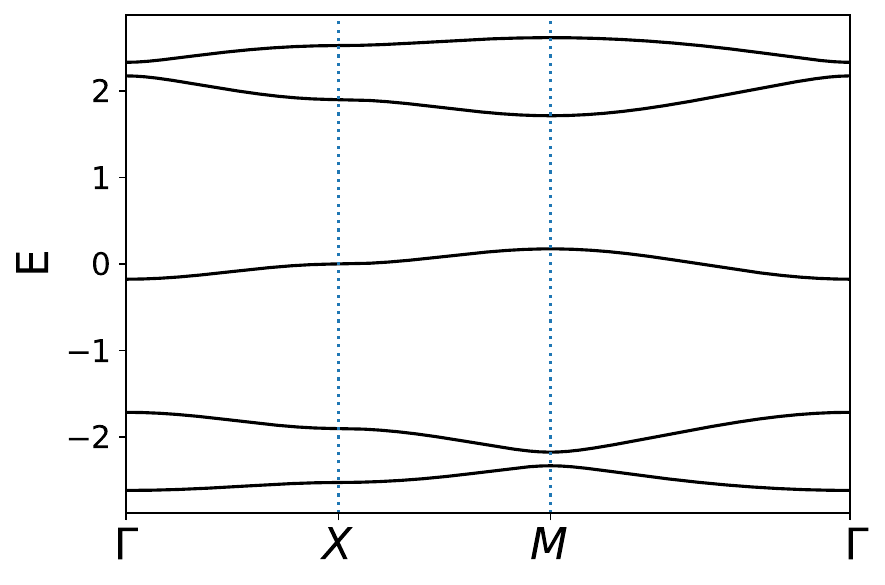}
}
\caption{Band spectra of the Hofstadter model for (a) $\varphi = 1/5$ and (b) $\varphi = 2/5$.}
\label{fig:Hofstadter_band_spectra}
\end{figure*}

The single particle band structures of the Hofstadter model are displayed in Fig.~\ref{fig:Hofstadter_band_spectra} along the high symmetry path $\Gamma (0,0) \to X(\pi/q, 0) \to M(\pi/q, \pi/q) \to \Gamma(0,0)$ of the magnetic wallpaper group $p4m'm'$.

\section{Static structure factors}\label{App:static_structure}

\begin{figure*}[ht]
\subfloat[]{
  \includegraphics[width=0.34\linewidth]{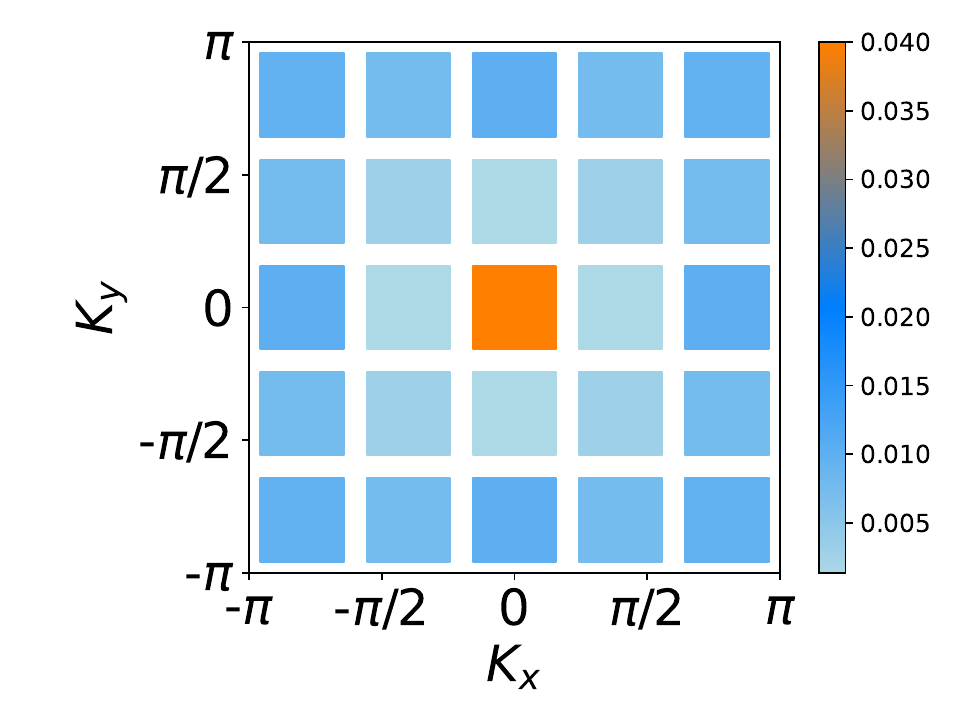}
}
\subfloat[]{
  \includegraphics[width=0.34\linewidth]{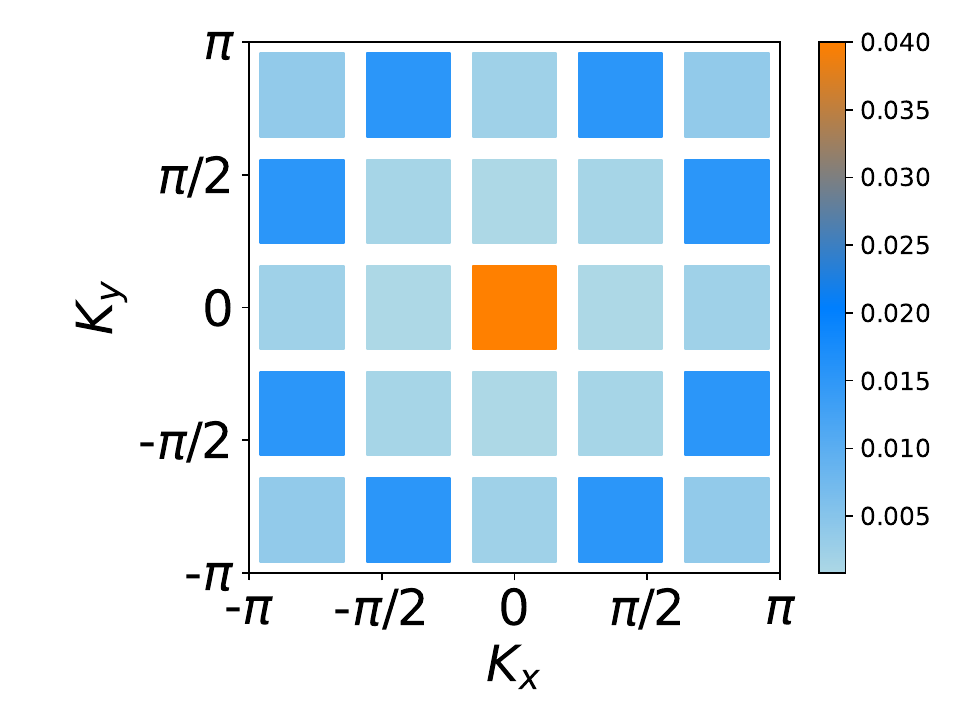}
}\hfil
\subfloat[]{
  \includegraphics[width=0.34\linewidth]{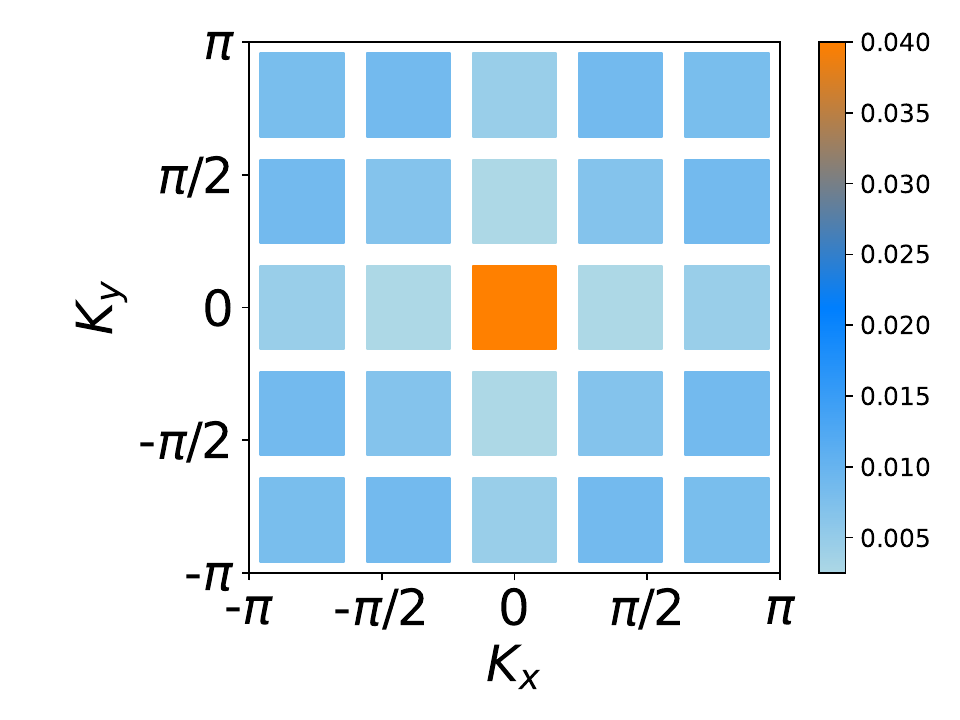}
}
\subfloat[]{
  \includegraphics[width=0.34\linewidth]{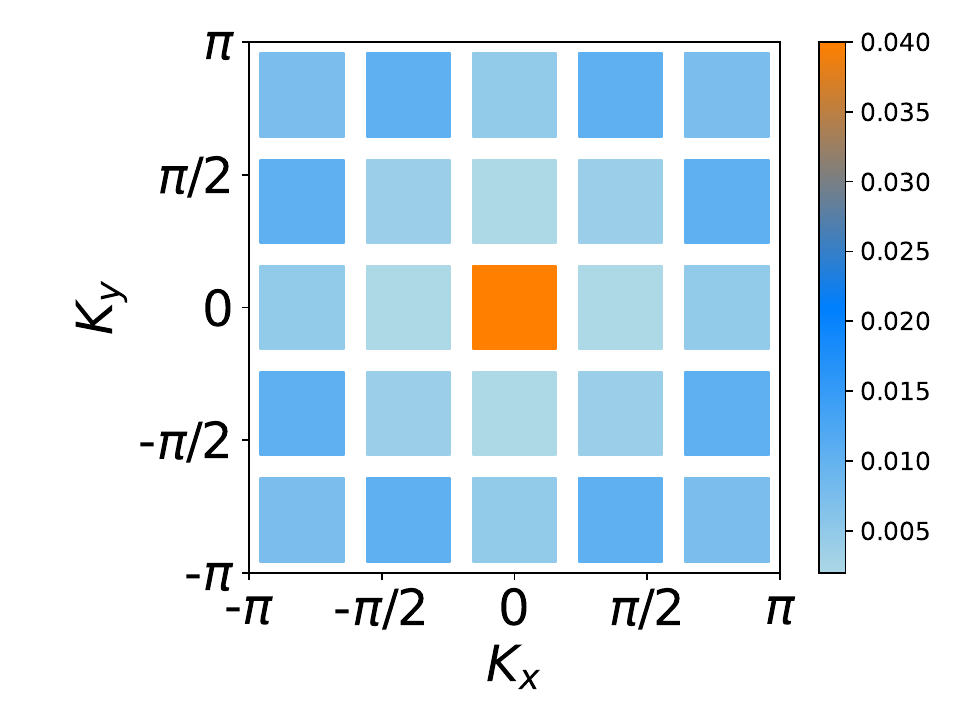}
}\hfil
\subfloat[]{
  \includegraphics[width=0.34\linewidth]{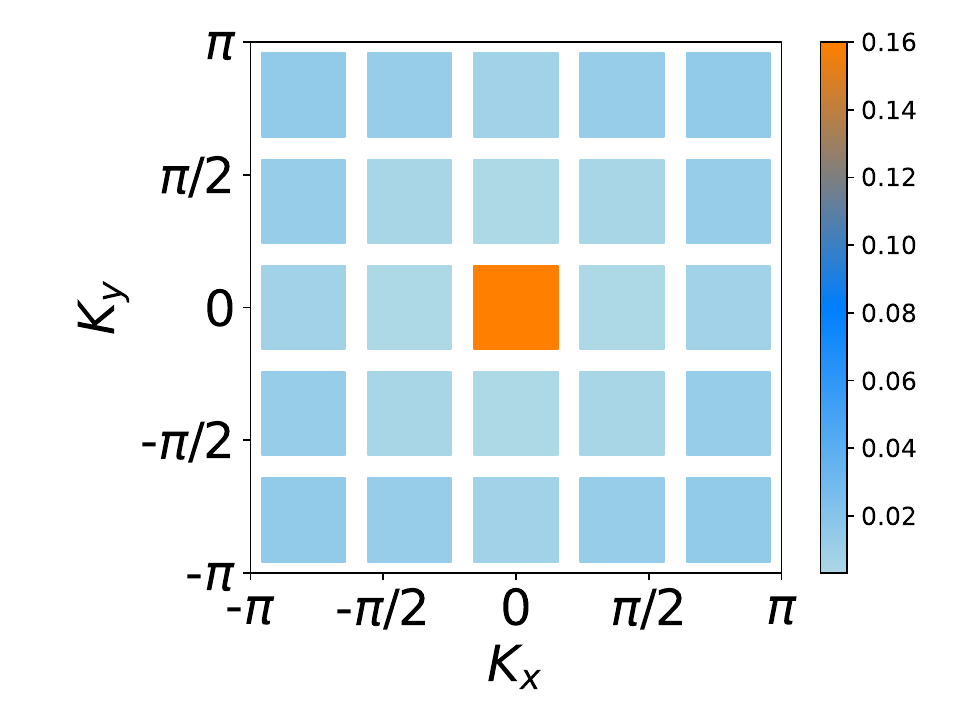}
}
\subfloat[]{
  \includegraphics[width=0.34\linewidth]{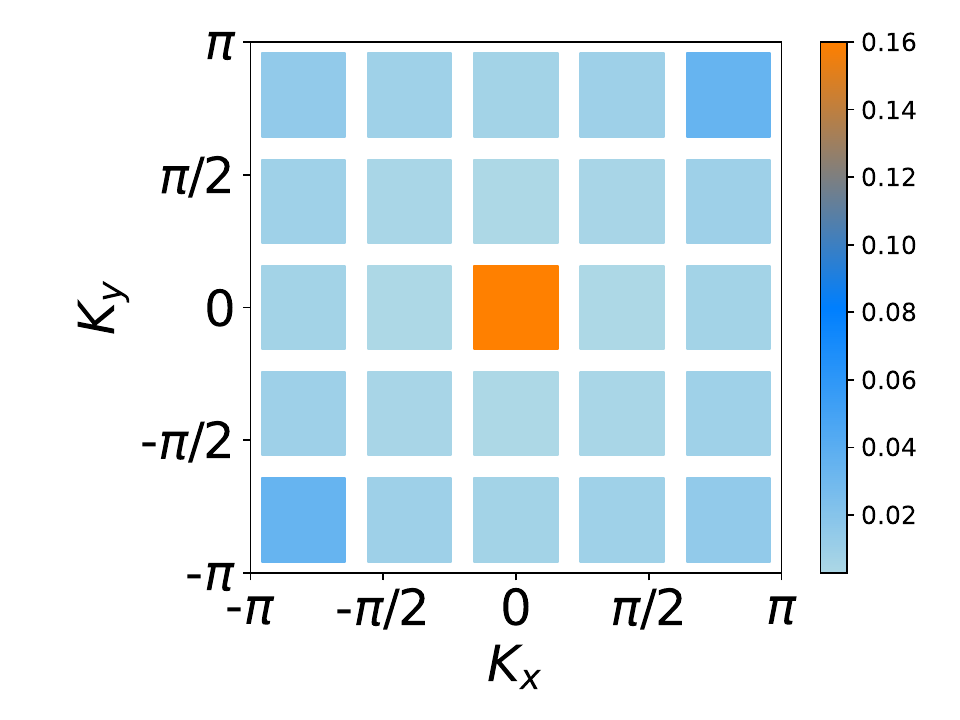}
}\hfil
\subfloat[]{
  \includegraphics[width=0.34\linewidth]{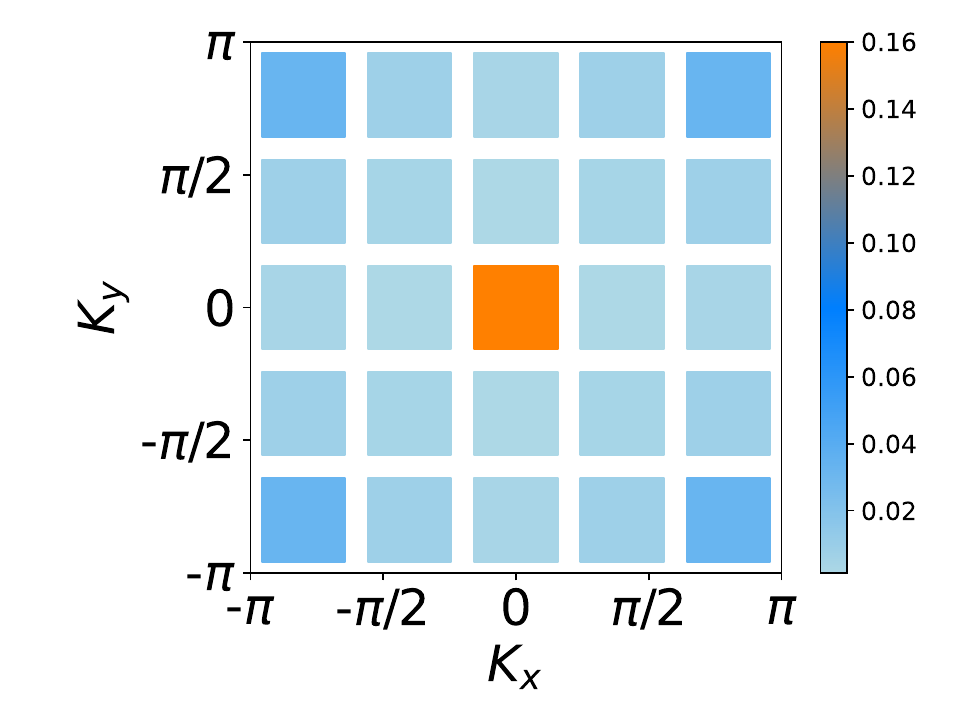}
}
\subfloat[]{
  \includegraphics[width=0.34\linewidth]{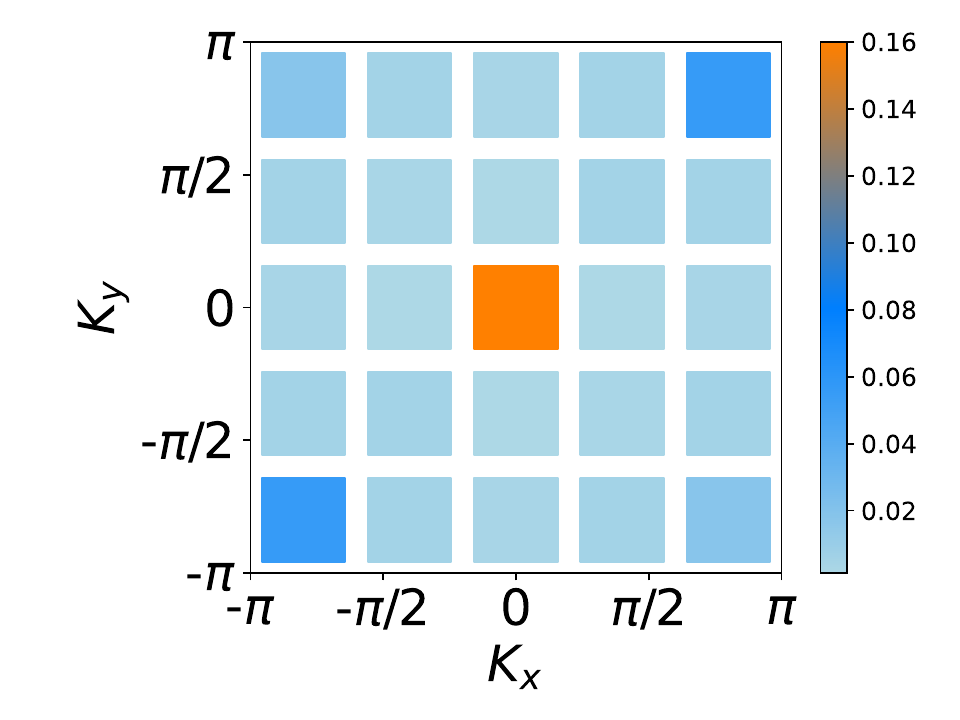}
}
\caption{Static structure factors of the Hofstadter Model on $5 \times 5$ clusters shortly before (left panels) and after (right panels) the phase transition. $N_e = 5$ and $\varphi= 1/5$ (a),(b), $N_e = 5$ and $\varphi= 2/5$  (c),(d), $N_e = 10$ and $\varphi= 1/5$ (e),(f), $N_e = 10$ and $\varphi= 2/5$ (g),(h).}
\label{fig:Hofstadter_structure_factors}
\end{figure*}

Static structure factors of the Hofstadter model for $\varphi, \rho = 1/5, 2/5$ are displayed in Fig.~\ref{fig:Hofstadter_structure_factors}.

\section{Ground States Under Flux Insertion}\label{App:first_appendix}

In this appendix we want to deduce Eq.~(\ref{Eq:flux_operators_algebra}) in the fractional quantum Hall regime and discuss the consequences for the CDW phase.

On a torus, a FQH phase has a $D$-fold quasi-degenerate ground state. Let the particle density be
\begin{equation}
    \rho = \frac{n}{q D},
\end{equation}
where $n$ and $D$ are coprime. We consider in the following only non-LSM clusters with $N_x = n_x q$ and $N_y = n_y q$. The algebra of magnetic translation operators obeys
\begin{equation}\label{Eq:fractional_translation}
    \hat{T}_{\vec{x}}^{M} \hat{T}_{\vec{y}}^{M} = \e^{-2\pi \varphi \ci N_e} \hat{T}_{\vec{y}}^{M} \hat{T}_{\vec{x}}^{M},
\end{equation}
where
\begin{equation}\label{Eq:fractional_translation_phase_factor}
    \e^{-2\pi \varphi \ci N_e} = \e^{-2\pi \frac{p}{q} \ci N_x N_y \rho} = \e^{-2\pi \ci \frac{p n n_x n_y}{D}} = \e^{-2\pi \ci \frac{m}{D}}
\end{equation}
and we stipulate coprime $m$ and $D$ (note that this may not always be possible depending on the flux quantum ratio and particle number per unit cell, e.g. in the case of $n = 1$, $q = 7$ and $D = 3$). As a consequence, the ground states are exactly degenerate. Under continuous flux insertion of a flux $\Phi$ during a time $T$, say, without loss of generality in $\vec{y}$-direction, the hoppings of the Hamiltonian change as
\begin{equation}\label{Eq:continuous_flux_insertion}
    c_{m+1, n}^{\dagger} c_{m,n} \to c_{m+1, n}^{\dagger} c_{m,n} \e^{-\ci \frac{\Phi}{\Phi_0} \frac{1}{N_x} \frac{t}{T}},
\end{equation}
which is unitarily equivalent to a twisted boundary of $\Phi t/\Phi_0 T$. The form of Eq.~(\ref{Eq:continuous_flux_insertion}) has the advantage that it leaves the Hamiltonian translation invariant, i.e.
\begin{equation}\label{Eq:combined_flux_insertion_gauge_trafo}
    [\hat{T}_{\alpha}^{M}, \hat{H}(\Phi t/T)] = 0,
\end{equation}
where here and in the following $\alpha$ will denote either $x$ or $y$. Note that the insertion of a flux quantum $\Phi_0$ does not produce a measurable Aharonov-Bohm phase, as it is periodic in $\Phi_0$. Consequently, $\hat{H}(\Phi_0)$ and $\hat{H}(0)$ are equivalent up to a gauge transformation.

We define the so-called ``large gauge transformation''~\cite{Lu}
\begin{equation}\label{Eq:large_gauge_transformation}
    \hat{U}_{\alpha} = \exp \left( \frac{-2\pi \ci}{L_{\alpha}} \displaystyle\sum_{m,n} r_{\alpha} c_{m,n}^{\dagger} c_{m,n} \right),
\end{equation}
where $r_{\alpha}$ is the $\alpha$-component of $(m,n)$. It is easy to verify that $\hat{H}(\Phi_0) = \hat{U}_{\alpha} \hat{H}(0) \hat{U}_{\alpha}^{\dagger}$. We can now define the operator
\begin{equation}
    \tilde{\hat{F}}_{\alpha}(\Phi_0) = \hat{U}_{\alpha}^{\dagger} \hat{F}_{\alpha}(\Phi_0)
\end{equation}
with the usual flux insertion operator
\begin{equation}
    \hat{F}_{\alpha}(\Phi) = \hat{T} \exp \left( -\frac{\ci}{\hbar} \int_{0}^{T} \diff t \hat{H}(\Phi t/T)  \right)
\end{equation}
 and time ordering operator $\hat{T}$. 
Consequently, $\tilde{\hat{F}}_{\alpha}(\Phi_0)$ is to be interpreted  as an operator that introduces a flux quantum $\Phi_0$ adiabatically through the action of $\hat{F}_{\alpha}(\Phi_0)$ for large enough $T$, and then the gauge transformation with 
 $\hat{U}_{\alpha}^{\dagger}$ converts $\hat{H}(\Phi_0)$ to the original Hamiltonian $\hat{H}(0)$. Summing up, $\tilde{\hat{F}}_{\alpha}(\Phi_0)$ maps a ground state of the Hamiltonian to a ground state of the Hamiltonian in the same gauge.

In the presence of translation invariance we can say more about the algebra of $\tilde{\hat{F}}_{\alpha}(\Phi_0)$. If translations commute with the Hamiltonian, then any translation will commute with $\hat{F}_{\alpha}(\Phi_0)$; on the other hand, by defining $\bar{\alpha} = y$ if $\alpha = x$ and $\bar{\alpha} = x$ if $\alpha = y$
\begin{equation}
    \hat{T}_{\alpha}^{M} \hat{U}_{\alpha} = \hat{U}_{\alpha} \hat{T}_{\alpha}^{M} \exp \left( 2\pi \ci \rho N_{\bar{\alpha}} \right) = \hat{U}_{\alpha} \hat{T}_{\alpha}^{M} \exp \left( 2\pi \ci \frac{n n_{\bar{\alpha}}}{D} \right)
\end{equation}
with coprime $n n_{\bar{\alpha}}$ and $D$ according to Eq.~(\ref{Eq:fractional_translation_phase_factor}) and
\begin{equation}
    \hat{T}^{M}_{\alpha} U_{\bar{\alpha}} = U_{\bar{\alpha}} \hat{T}^{M}_{\alpha}.
\end{equation}
This in turn implies
\begin{equation}
    \hat{T}^{M}_{\alpha} \tilde{\hat{F}}_{\alpha}(\Phi_0) = \tilde{\hat{F}}_{\alpha}(\Phi_0) \hat{T}^{M}_{\alpha} \exp \left( -2\pi \ci \frac{n n_{\alpha}}{D} \right),
\end{equation}
whereas $\hat{T}^{M}_{\alpha}$ and $\tilde{\hat{F}}_{\bar{\alpha}}(\Phi_0)$ commute.

Now, let $\ket{\Psi}$ be an eigenstate of $T_{\vec{y}}^{M}$,
\begin{equation}
    \hat{T}_{\vec{y}}^{M} \ket{\Psi} = \e^{-\ci K_y} \ket{\Psi}.
\end{equation}
Then
\begin{equation}
    \hat{T}^{M}_{\vec{y}} (\hat{T}_{\vec{x}}^{M} \ket{\Psi}) = \e^{2\pi \ci \frac{m}{D}} \e^{-\ci K_y} (\hat{T}_{\vec{x}}^{M} \ket{\Psi}),
\end{equation}
but also
\begin{equation}
    \hat{T}^{M}_{\vec{y}} (\tilde{\hat{F}}_{\vec{y}}(\Phi_0)^{-p n_y} \ket{\Psi}) = \e^{2\pi \ci \frac{m}{D}} \e^{-\ci K_y} (\tilde{\hat{F}}_{\vec{y}}(\Phi_0)^{-p n_y} \ket{\Psi}).
\end{equation}
Since all $D$ states carry a unique $K_y$ momentum, and an adiabatic flux insertion can only map a ground state to another ground state, it follows that
\begin{equation}
    \tilde{\hat{F}}_{\vec{y}}(\Phi_0)^{-p n_y} \ket{\Psi} \propto \hat{T}_{\vec{x}}^{M} \ket{\Psi}
\end{equation}
and
\begin{equation}
    \tilde{\hat{F}}_{\vec{x}}(\Phi_0) \ket{\Psi} \propto \ket{\Psi}.
\end{equation}
Vice versa one finds in the eigenbasis of $\hat{T}_{\vec{x}}^{M}$
\begin{equation}
    \hat{T}^{M}_{\vec{x}} (\tilde{\hat{F}}_{\vec{x}}(\Phi_0)^{-p n_x} \ket{\Psi}) = \e^{2\pi \ci \frac{m}{D}} \e^{-\ci K_x} (\tilde{\hat{F}}_{\vec{x}}(\Phi_0)^{-p n_x} \ket{\Psi})
\end{equation}
and $\tilde{\hat{F}}_{\vec{x}}(\Phi_0)$ is now diagonal. This implies a similar transformation behavior between translations and flux insertion according to
\begin{equation}
    \begin{aligned}[b]
    \tilde{\hat{F}}_{\vec{y}}(\Phi_0)^{-p n_y} \cong & \hat{T}_{\vec{x}}^{M} \\
    \tilde{\hat{F}}_{\vec{x}}(\Phi_0)^{-p n_x} \cong & \left( \hat{T}_{\vec{y}}^{M} \right)^{\dagger}.
    \end{aligned}
\end{equation}
As a consequence we find for the commutation relation between the two flux insertion operators
\begin{equation}
    \tilde{\hat{F}}_{\vec{x}}(\Phi_0)^{p n_x} \tilde{\hat{F}}_{\vec{y}}(\Phi_0)^{p n_y} = \e^{-2\pi \ci \frac{m}{D}} \tilde{\hat{F}}_{\vec{y}}(\Phi_0)^{p n_y} \tilde{\hat{F}}_{\vec{x}}(\Phi_0)^{p n_x}
\end{equation}
or
\begin{equation}
    \tilde{\hat{F}}_{\vec{x}}(\Phi_0) \tilde{\hat{F}}_{\vec{y}}(\Phi_0) = \e^{2\pi \ci \frac{nf}{D}} \tilde{\hat{F}}_{\vec{y}}(\Phi_0) \tilde{\hat{F}}_{\vec{x}}(\Phi_0),
\end{equation}
where $f$ is some natural number coprime to $D$, so that $fp = 1 \text{ mod } D$, which exists according to B\'ezout's identity.

How do the above results change in a CDW phase? In that case, the symmetry broken states should be eigenstates of flux insertion operations as argued in Sec.~\ref{subsec:first_subsec}, hence
\begin{equation}
    \tilde{\hat{F}}_{\alpha}(\Phi_0) \ket{\Psi} = \e^{\ci \gamma_{\alpha}} \ket{\Psi}.
\end{equation}
We can transform $\ket{\Psi}$ into a state orthogonal to it by translation:
\begin{equation}
\begin{aligned}[b]
    & \tilde{\hat{F}}_{\alpha}(\Phi_0) \left( \hat{T}_{\vec{x}}^{M} \right)^{l_x} \left( \hat{T}_{\vec{y}}^{M} \right)^{l_y} \ket{\Psi} \\
    = & \exp \left( 2\pi \ci \frac{n n_{\bar{\alpha}} l_{\alpha}}{D} \right) \e^{\ci\gamma_{\alpha}} \left( \hat{T}_{\vec{x}}^{M} \right)^{l_x} \left( \hat{T}_{\vec{y}}^{M} \right)^{l_y} \ket{\Psi}
\end{aligned}
\end{equation}
Therefore, there must be at least $D^2$ orthogonal quasi-degenerate CDW states. To be more precise, if
\begin{equation}
    \left( \hat{T}_{\alpha}^{M} \right)^{D} \ket{\Psi} \propto \ket{\Psi},
\end{equation}
then the CDW states transform exactly like single particles in a $D \times D$ cluster with periodic boundary conditions and $m/D$ flux quanta per unit cell. It is well known that these states belong to $D_{M}^{\vec{K} = 0}$, the irreducible projective representation of the translation group at $\vec{K} = 0$. In general $\vec{K}$ distinguishes all different possible projective representations of the translation group belonging to a cluster, just like the well-known one-dimensional irreducible representations of the translation group~\cite{Bro}. Assume
\begin{equation}
    \left( \hat{T}_{\alpha}^{M} \right)^{D \delta_{\alpha}} \ket{\Psi} \propto \ket{\Psi}
\end{equation}
with the smallest possible $\delta_{\alpha}$, then all of the CDW states belong to $D$ copies of $D_{M}^{\vec{K}}$ for each $K_{\alpha} = 0, 2\pi / \delta_{\alpha}, \dots 2\pi (\delta_{\alpha} - 1)/\delta_{\alpha}$. Finally, we want to make two remarks. In the case $\delta_{\alpha} = 1$ (this could for example be realized for $q = D = 3$), all CDW states transform under the same projective irrep $D_{M}^{\vec{K} = 0}$ which does not contradict the notion of symmetry breaking, as the projective irreducible representations of the translation group are higher dimensional. Secondly, the minimal required number of quasi-degenerate ground states in the CDW regime is higher than would be the case for ordinary translation symmetry (without an external magnetic field), where both the CDW and the topologically ordered phase can have the same number of ground states, providing no clear sign for topological order.

\bibliography{main}

\end{document}